\documentclass[final,onefignum,onetabnum]{siamart190516}
\usepackage{soul} 

\usepackage{xcolor,cite}

\usepackage{lipsum}
\usepackage{amsfonts}
\usepackage{graphicx}
\usepackage{epstopdf}
\usepackage{algorithmic}
\ifpdf
  \DeclareGraphicsExtensions{.eps,.pdf,.png,.jpg}
\else
  \DeclareGraphicsExtensions{.eps}
\fi

\newsiamremark{remark}{Remark}
\newsiamremark{hypothesis}{Hypothesis}
\crefname{hypothesis}{Hypothesis}{Hypotheses}
\newsiamthm{claim}{Claim}

\headers{Distinct excitatory/inhibitory bump wandering}{H.L. Cihak, T.L. Eissa, and Z.P. Kilpatrick}

\title{Distinct excitatory and inhibitory bump wandering in a stochastic neural field \thanks{Submitted to the editors DATE.
\funding{This work was funded by National Institutes of Health grants R01MH115557-01 and R01-EB029847-01 as well as an NSF grant DMS-1853630.}}}

\author{Heather L Cihak\thanks{Department of Applied Mathematics, University of Colorado Boulder, Boulder, CO 
  (\email{Heather.Cihak@colorado.edu}, \email{Tahra.Eissa@colorado.edu}, \email{zpkilpat@colorado.edu}}
\and Tahra L Eissa\footnotemark[2] \and Zachary P Kilpatrick\footnotemark[2]
}

\usepackage{amsopn}

\usepackage{xr}
\makeatletter
\newcommand*{\addFileDependency}[1]{
  \typeout{(#1)}
  \@addtofilelist{#1}
  \IfFileExists{#1}{}{\typeout{No file #1.}}
}
\makeatother

\newcommand*{\myexternaldocument}[1]{%
    \externaldocument{#1}%
    \addFileDependency{#1.tex}%
    \addFileDependency{#1.aux}%
}

\ifpdf
\hypersetup{
  pdftitle={Distinct excitatory and inhibitory bump wandering in a stochastic neural field},
  pdfauthor={H. Cihak, T.E. Eissa, and Z.P. Kilpatrick}
}
\fi

\myexternaldocument{ex_supplement}

\begin{document}

\maketitle
\begin{abstract}
  Localized persistent cortical neural activity is a validated neural substrate of parametric working memory. Such activity `bumps' represent the continuous location of a cue over several seconds. Pyramidal (excitatory) and interneuronal (inhibitory) subpopulations exhibit tuned bumps of activity, linking neural dynamics to behavioral inaccuracies observed in memory recall.  However, many bump attractor models collapse these subpopulations into a single joint excitatory/inhibitory (lateral inhibitory) population, and do not consider the role of interpopulation neural architecture and noise correlations. Both factors have a high potential to impinge upon the stochastic dynamics of these bumps, ultimately shaping behavioral response variance. In our study, we consider a neural field model with separate excitatory/inhibitory (E/I) populations and leverage asymptotic analysis to derive a nonlinear Langevin system describing E/I bump interactions. While the E bump attracts the I bump, the I bump stabilizes but can also repel the E bump, which can result in prolonged relaxation dynamics when both bumps are perturbed. Furthermore, the structure of noise correlations within and between subpopulations strongly shapes the variance in bump position. Surprisingly, higher interpopulation correlations reduce variance.
\end{abstract}




\section{Introduction}
\label{sec:Introduction}
Storing information `in mind’ for short periods of time is essential for performance of daily tasks~\cite{GoldmanRakic1995}. Parametric working memory (as used in tasks requiring a delayed estimate of a continuous quantity) uses spatially localized persistent neural activity in the prefrontal cortex, parietal cortex, and frontal eye fields~\cite{Curtis06, GoldmanRakic1995} generated in neural circuits with strong local recurrent excitation and broad inhibition~\cite{Constantinidis16, Kilpatrick13WandBumpSIAD, Seeholzer19}.
Neurophysiological recordings from non-human primate subjects performing visuospatial working memory tasks have shown that localized bumps of persistent activity encode remembered parametric quantities for a few seconds~\cite{Constantinidis16, Kilpatrick13WandBumpSIAD,Funahashi89,GoldmanRakic1995}. For example, in the oculomotor delayed response task, a location cue is momentarily presented on a circle displayed on a monitor, then a {\em delay period} occurs during which the video is blank, and finally the subject is prompted to report their memory of the cued location. During the delay period, neural recordings identify cells tuned to specific cue locations around the circle and reveal that the strongest (peak) firing neurons represent the encoded location ~\cite{GoldmanRakic1995,Funahashi89,Fuster73}.
Peaked and localized activity wanders stochastically during the delay period, representing a time-dependent degradation of cue location memory consistent with subsequent response errors~\cite{Compte2000,Wimmer14}.

Neural field and spiking models are capable of producing peaked, localized activity that wanders via spatially structured connectivity that weakens as the distance between neural cue location preference increases with excitation having a narrower spatial footprint~\cite{WilsonCowan73,Amari77,Seeholzer19, Kilpatrick13WandBumpSIAD}.
Bumps can be defined as standing pulse solutions usually with two key dynamical properties important for encoding memories of continuum variables: (1) they are self-sustained in the absence of stimulus (representing memory over a delay period) and (2) they exhibit marginal stability,  such that they can occur at any location in the network and integrate translational perturbations~\cite{Amari77}. Such solutions have garnered interest due to the rich dynamical features that arise when varying the form of connectivity, introducing propagation delays, adding slow negative feedback, or considering stochasticity~\cite{Blomquist05,Amari77,Kilpatrick13WandBumpSIAD,coombes2003waves,folias2004breathing}. Response statistics and neural activity during oculomotor delayed response tasks are also well characterized by the output of bump attractor models with some form of stochasticity incorporated~\cite{Wimmer14,Constantinidis16,barbosa2020interplay}.

Previous psychophysical studies of delayed estimation of continuous quantities show subject errors scale linearly with the delay period ~\cite{Ploner98,White94}. Such behavior can be accounted for by models whose low-dimensional dynamics evolve as Brownian motion, like a particle subject to diffusion. This is consistent with a bump attractor stochastically perturbed by noise, wandering along a marginally stable ring attractor~\cite{Kilpatrick13WandBumpSIAD,Compte2000}. Extended models have also considered neural circuit mechanisms that help stabilize the movement of bumps to noise perturbations by breaking the marginal stability of the ring attractor. For example, spatially heterogeneous recurrent excitation leads to low-dimensional dynamics akin to a particle stochastically perturbed along a washboard potential, slowing the rate of diffusion~\cite{kilpatrick2013optimizing}. Likewise, short-term facilitation locally potentiates synaptic excitation where the bump is instantiated, akin to a slowly moving local potential well that traps the particle near the true memorandum location~\cite{Seeholzer19,Kilpatrick18STP_FacSCIREP}. In addition to stabilizing bumps within trials, short-term facilitation also transfers memory of the previous trial stimulus to the next, creating systematic errors referred to as serial dependence~\cite{kiyonaga2017serial,cicchini2018functional}.

Despite advancements in understanding how these modifications to bump attractor models affect their stochastic dynamics, a detailed examination of the role of separate excitatory and inhibitory (E and I) population dynamics has been overlooked, and I-I interactions are often ignored entirely. Inhibition is often assumed to be flat and global in stochastic bump attractor models~\cite{Constantinidis16}, but we know prefrontal cortical synaptic inhibition exhibits nontrivial preference-dependent interactions~\cite{GoldmanRakic1995}, which could have yet unidentified effects on the dynamics of bumps and the fidelity of memory systems that rely on them.

Here, we  use a stochastic neural field model to investigate the wandering dynamics and interactions of separate E and I activity bumps. We focus on how a separate and spatially-extended I population shapes the variance of a bump attractor encoding a remembered location along a continuum. First, we introduce the neural field equations, notation conventions, and parameters whose effects on the variance of the bumps' final position will be analyzed (\Cref{sec:TheModels}). We then review the existence and stability of both E and I bump solutions~\cite{WilsonCowan73,Amari77,Blomquist05,Coombes2012InterfaceJNeuro,Laing2001,FoliasErmentrout2011}; paving the way for examining how deterministic and stochastic perturbations deform and shift the attractor solution. We examine the impact of modifying both the amplitude and spatial extent of neural architecture, as well as the firing rate thresholds on the stability of solutions (\Cref{sec:Deterministic}). Stable and unstable branches of bumps annihilate at a discontinuous saddle-node bifurcation given a sufficiently high firing rate thresholds.

Our stochastic analysis involves two different methods of identifying the impact of noise on the bump attractor, (1) a linear perturbative analysis (Strongly Coupled Limit between the E and I bumps) in the limit of weak noise and (2) an Interface Based analysis where we track the threshold crossing points of the E/I bumps (\Cref{sec:Stochastic}). Thus, we develop two corresponding theoretical predictions of bump position variance, the second of which captures nonlinear interactions between the E/I bumps in a Langevin system that is a better match to full model simulations. This higher-order analysis starts by obtaining distinct and nonlinearly coupled equations for the bump interfaces. Bump position variance changes non-monotonically as a function of most model parameters, so intermediate tuning of network architecture maximizes memory degradation. Finally, we identify the impact of noise correlations between the E and I populations; finding they actually serve to reduce bump position variance.


\section{The Model}
\label{sec:TheModels}
\begin{figure}[htbp]
  \centering
  \includegraphics[width=\textwidth]{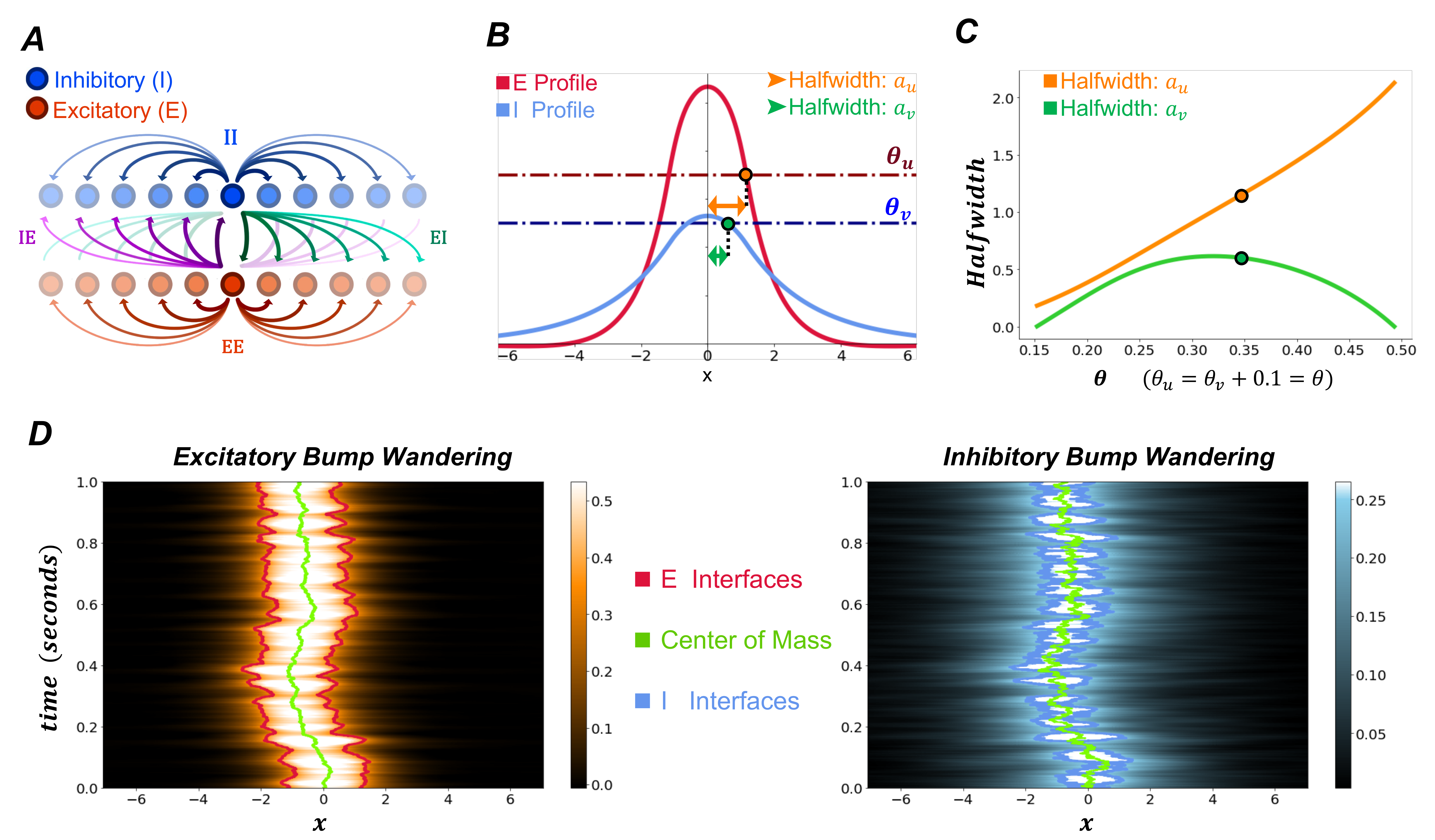}
  \caption{\textbf{Independence of Excitatory and Inhibitory bumps.} (A) Model schematic with separate E and I populations and the synaptic connections between them. (B) An example pair of E and I bump profiles  with firing rate thresholds $\theta_u=0.35$ and  $\theta_v=0.25$ and $A_{ii}=0$. 
  (C) Example plot of solution half-widths as thresholds are varied, obtained by numerically solving the threshold conditions, Eq.~(\ref{eq:3.4}), for $a_u$ and $a_v$. The particular half-width solution from panel B is identified by the corresponding dots. 
  (D) Example bumps from panel B wandering over time with noise amplitude $\epsilon=0.002$. Traces represent the center of mass (green) and the E/I interfaces (red/blue, respectively). Colorscales represent the profile amplitude of $u(x,t)$ and $v(x,t)$. Other parameters are as in \cref{tableparam}. }
  \label{fig:1}
\end{figure}

To determine the effects of separately evolving excitatory (E) and inhibitory (I) populations on bump attractor dynamics, we analyze a stochastic neural field model: a coupled system of nonlinear integro-differential equations describing interactions of spatially-extended E and I neural sub-populations~(\Cref{fig:1}A and \cite{WilsonCowan73}). Recurrent connectivity targeting position $x$ from $y$ at time $t$ is described by convolving ($a(x)*b(x) = \int_{- \infty}^{\infty} a(x-y) b(y) dy$) the synaptic kernel and the firing rate output of the neurons from which synapses originate.
Along with additive noise terms, we obtain the stochastic neural field equations: 
\begin{subequations}
\begin{align}
    du(x,t)=\left[-u(x,t)+{w_{ee}(x)*f(u(x,t))}-{w_{ei}(x)*f(v(x,t))}\right]dt+\epsilon^{\frac{1}{2}} dW_{e}\\
    \tau dv(x,t)=\left[-v(x,t)+{w_{ie}(x)*f(u(x,t))}-{w_{ii}(x)*f(v(x,t))}\right]dt+\epsilon^{\frac{1}{2}}dW_{i}.
\end{align}\label{eq:2.1}
\end{subequations}
where $u(x,t)$ and $v(x,t)$ denote the E and I synaptic input profiles at location $x$ at time $t$. We drop the time constant from the E population and take time units to be 10ms, on the order of typical E membrane and synaptic time constants~\cite{hausser1997estimating}. The sigmoidal firing rate function $f(u)=\frac{1}{1+e^{-\eta (u-\theta)}}$ with threshold $\theta$ and gain $\eta$ determines the strength of the firing rate output based on the input $u$. Analytical results can be obtained in the high gain limit ($\eta \to \infty$) such that
\begin{align}
    f(u) = H(u-\theta)=\begin{cases}1 & u-\theta \geq 0,\\ 0 & u-\theta < 0,\end{cases}\label{eq:2.2}
\end{align}
is the Heaviside function.  The E and I population firing rate thresholds, $\theta_u$ and $\theta_v$, can differ. Given a Heaviside firing rate nonlinearity, it is possible to exactly characterize the dynamics of the model using {\em interface} equations that track the evolution of level sets $u = \theta_u$ and $v = \theta_v$ in space and time~\cite{Coombes2012InterfaceJNeuro,Faye2018}. The strength of synaptic connectivity weakens with the spatial distance between positions $x$ and $y$ and is given by the distance-dependent synaptic weight profile functions $w_{ab}(x-y)$ ($a,b \in \{ e, i\}$). For explicit calculations, we take these to be symmetric exponentials  
\begin{align}
    w_{ab}(x)=A_{ab}e^{-\frac{|x|}{\sigma_{ab}}}, \label{expwt}
\end{align}
where $a,b\in\{e,i\}$ and $A_{ab},\sigma_{ab}\in \mathbb{R}^{\geq 0}$ (non-negative constants). Commonly used values throughout for the weight profiles are presented in \cref{tab:params}. Parameters for $w_{ee}$ were chosen to non-dimensionalize the E to E connectivity (setting $A_{ee} =0.5$ and $\sigma_{ee} = 1$). Other parameters for the I to I and cross population synaptic profiles were generally chosen to be broader than the synaptic footprint of E to E connections, and we generally assume the commonly identified $80\%$ E and $20\%$ I neuron ratio~\cite{abeles1991corticonics} in turn approximately determines the effective synaptic amplitude from those populations.

\begin{table}[t!]
\label{tableparam}
{\footnotesize
  \caption{\textbf{Model parameters for Eq.~\cref{eq:2.1}}}\label{tab:params}
\begin{center}
  \begin{tabular}{|c|p{11mm}|p{11mm}|p{11mm}|p{11mm}|p{20mm}|p{12mm}|} \hline
   \bf{Parameter} & $A_{ee}$ & $A_{ei},A_{ie}$ & $A_{ii}$ & $\sigma_{ee}$ & $\sigma_{ei},\sigma_{ie},\sigma_{ii}$ &  $\tau$  \\ \hline
   \bf{Definition} & E-E strength & I-E strength & I-I strength & E-E spatial scale & other spatial scales & I time constant \\ \hline \bf{Value} & 0.5 & 0.15 & 0 or 0.01 & 1 & 2  & 1  \\ \hline
  \end{tabular}
\end{center}
}
\end{table}

The spatially-structured multiplicative noise terms
$dW_{e}=\sqrt{|u(x,t)|}dW_u(x,t)$ and $dW_{i}=\sqrt{|v(x,t)|}dW_v(x,t)$ are are introduced with small amplitude $0 < \epsilon \ll 1$, allowing asymptotic approximations of their stochastic effects. Multiplicative noise is predicted in neural fields by linearly approximating noise arising in a system-size expansion of a fully stochastic model~\cite{Bressloff2010}. Increments of spatially-extended Wiener processes have zero mean $\langle dW_a(x,t)\rangle=0$ ($a \in \{u, v\}$) and local spatial correlations $\langle dW_a(x,t)dW_a(y,s)\rangle=C_{aa}(x-y)\delta(t-s)dtds$. For simplicity, we start by assuming there are no correlations between the noise to the E and I populations $\langle d W_u(x,t) dW_v(y,s) \rangle \equiv 0 $.

Based on prior studies of deterministic E/I population models~\cite{Pinto2001b,Blomquist05,Compte2000} bump (standing pulse) solutions emerge in parameter regions we can identify using an existence/stability analysis. Recurrent excitation sustains both the E and I populations and inhibition prevents the spread of the E population activity. In the absence of noise, we obtain bump profiles that are even symmetric and translation symmetric. Using threshold-crossing conditions, we develop implicit formulas for the half-width variable $a_u$ ($a_v$) for the E (I) bump, which ensure existence, defined as the distance from the bump's center of mass to the interface (\cref{fig:1}B). 
The I bump's half-width $a_v$ (\cref{fig:1}C) can vary non-monotonically along certain parameter axes, such as increasing firing rate thresholds, $\theta_u$ and $\theta_v$, simultaneously.
When noise is applied to stable E/I activity bumps, they `wander' due to their neutral stability. In addition to small deformations of the bump profile itself, a bump's center of mass wanders stochastically (\cref{fig:1}D and see \cref{fig:EIWANDER} for an additional example). While past asymptotic analyses assumed this wandering was roughly Brownian motion~\cite{Compte2000,Kilpatrick13WandBumpSIAD}, we will show that by relaxing this assumption we can better characterize nonlinear interactions between the bumps in the distinct E and I populations, and how this shapes wandering dynamics.

Each bump has a region over which neural activity ($u$ or $v$) is superthreshold (above $\theta_u$ or $\theta_v$). We define the corresponding E/I population {\em active} regions to be $[x_1(t),x_2(t)]$ and $[x_3(t),x_4(t)]$ respectively~\cite{Amari77}. These active regions are bounded by the interfaces (threshold crossings), since 
$u(x_{1,2}(t),t)=\theta_u$ and $v(x_{3,4}(t),t)=\theta_v$. Note, large deviations could lead to multiple disjoint active region segments in each layers, but we assume each bump's active region is fully connected here (See \cite{Coombes2012InterfaceJNeuro,Faye2018,Krishnan18} for elaborations on this problem). In the absence of noise, the interfaces relate directly to the half-widths in that $a_u=(x_2-x_1)/2$ and $a_v=(x_4-x_3)/2$.

Stochastic motion of the bumps will be tracked by estimating the center of mass $ \Delta_u(t)$ ($ \Delta_v(t)$) of the active region of the E (I) bump
\begin{align}
    \Delta_u(t)= \frac{x_1(t)+x_2(t)}{2}, \hspace{2cm}
    \Delta_v(t)= \frac{x_3(t)+x_4(t)}{2}.\label{eq:2.3}
\end{align}
Overall, we are interested both in how network parameters shape the form and stability of bumps as well as how this translates into bump's stochastic dynamics in the presence of noise. Our ensuing analysis will initially rely on techniques in local stability as well as weak perturbations, but a major advancement will be the use of interface techniques to provide higher-order corrections to the effective nonlinear equations describing stochastic bump motion.

\section{Deterministic Analysis}
\label{sec:Deterministic}
Our estimates for expected bump wandering (variance of center of mass) utilizes linearization about stable stationary solutions. This necessitates an analysis of stationary bump solutions' (static bump solutions in the absence of perturbations) existence and stability in varying parameter regimes for the deterministic system.
Using Eq.~\cref{eq:2.1} without noise~\cite{Blomquist05,Amari77}, we demonstrate the existence of stationary solutions through direct construction. For proofs of uniqueness of solutions with a Heaviside firing rate see \cite{Blomquist05,NeuralFields2014}. 

\subsection{Stationary Solutions}
\label{sec:stationary}
Assuming solutions are stationary ($u(x,t)=U(x)$ and $v(x,t)=V(x)$) and using the Heaviside firing rate function Eq.~\cref{eq:2.2}, Eq.~\cref{eq:2.1} becomes
\begin{subequations}
\begin{align}
    & U(x)=\int_{\mathbb{R}}w_{ee}(x-y)H(U(y)-\theta_u)dy-\int_{\mathbb{R}}w_{ei}(x-y)H(V(y)-\theta_v)dy\\
    & V(x)=\int_{\mathbb{R}}w_{ie}(x-y)H(U(y)-\theta_u)dy-\int_{\mathbb{R}}w_{ii}(x-y)H(V(y)-\theta_v)dy
\end{align}\label{eq:3.1}
\end{subequations}
We seek bumps with simply connected active regions ($U>\theta_u$ and $V>\theta_v$) $[-a_u, a_u]$ and $[-a_v, a_v]$. Exploiting the solutions' translational invariance along $\mathbb{R}$ (opting for a center of mass at $x=0$) and expected evenness (i.e. $U(-x)=U(x)$ and $V(-x)=V(x)$) we obtain the system
\begin{subequations}
\begin{align}
    & U(x)=\int_{-a_u}^{a_u}w_{ee}(x-y)dy-\int_{-a_v}^{a_v}w_{ei}(x-y)dy\\
    & V(x)=\int_{-a_u}^{a_u}w_{ie}(x-y)dy-\int_{-a_v}^{a_v}w_{ii}(x-y)dy
\end{align}\label{eq:3.2}
\end{subequations}

Integrating our chosen exponential synaptic weight functions ($w_{ab}(x)=A_{ab}e^{\frac{-|x|}{\sigma_{ab}}}$, $a,b\in\{e,i\}$), we find the explicit formulas for $c \in \{ a_u, a_v \}$:
\begin{align}\int^c_{-c}w_{ab}(x-y)dy=\begin{cases}2A_{ab}\sigma_{ab}e^{\frac{-x}{\sigma_{ab}}}\sinh\left(\frac{c}{\sigma_{ab}}\right)& x>c,\\
2A_{ab}\sigma_{ab}\left[1-e^{\frac{-c}{\sigma_{ab}}}\cosh\left(\frac{x}{\sigma_{ab}}\right)\right]& |x|<c,\\
2A_{ab}\sigma_{ab}e^{\frac{x}{\sigma_{ab}}}\sinh\left(\frac{c}{\sigma_{ab}}\right)& x<-c.
\end{cases}\label{eq:3.3}\end{align}

\noindent Substituting Eq.~\cref{eq:3.3} into Eq.~\cref{eq:3.2} and utilizing the threshold crossing conditions $\theta_u=U(\pm a_u)$ and $\theta_v=V(\pm a_v)$, we obtain an implicit set of equations for the half-widths, which depends on whether the E or I bump is wider:
\begin{subequations}
\begin{align}
    &\theta_u=2A_{ee}\sigma_{ee}e^{\frac{-a_u}{\sigma_{ee}}}\sinh\left(\frac{a_u}{\sigma_{ee}}\right)-\begin{cases}2A_{ei}\sigma_{ei}e^{\frac{-a_u}{\sigma_{ei}}}\sinh\left(\frac{a_v}{\sigma_{ei}}\right)& a_u\geq a_v,\\
    2A_{ei}\sigma_{ei}\left[1-e^{\frac{-a_v}{\sigma_{ei}}}\cosh\left(\frac{a_u}{\sigma_{ei}}\right)\right]& a_u<a_v,
    \end{cases}\\
    &\theta_v=-2A_{ii}\sigma_{ii}e^{\frac{-a_v}{\sigma_{ii}}}\sinh\left(\frac{a_v}{\sigma_{ii}}\right)+\begin{cases}2A_{ie}\sigma_{ie}e^{\frac{-a_v}{\sigma_{ie}}}\sinh\left(\frac{a_u}{\sigma_{ie}}\right)& a_v\geq a_u,\\
    2A_{ie}\sigma_{ie}\left[1-e^{\frac{-a_u}{\sigma_{ie}}}\cosh\left(\frac{a_v}{\sigma_{ie}}\right)\right]& a_v<a_u.
    \end{cases}
\end{align}\label{eq:3.4}
\end{subequations}
Eq.~\cref{eq:3.4} is thus defined piecewise continuously (\cref{fig:2}A, gold circle indicates the change in cases occurs) 
and, as we will show, cusps appear at these case switches in plots of eigenvalues and variance estimates.

Note, the above set of threshold conditions assume that both the E and I population have nontrivial active regions. There is a second class of bump solutions where there is no active region in the I population. As we will show, these bumps always tend to be linearly unstable as there is no active inhibition to prevent the spread of excitation upon perturbation. Applying this assumption to Eq.~\cref{eq:3.2}, we obtain the simplified system,
\begin{subequations}
\begin{align}
    & U(x)=\int_{-a_u}^{a_u}w_{ee}(x-y)dy,\\
    & V(x)=\int_{-a_u}^{a_u}w_{ie}(x-y)dy,
\end{align}\label{eq:3.5}
\end{subequations}
with E bump half-width given by
\begin{equation*}
    \theta_u=U(\pm a_u)=\int_{-a_u}^{a_u}w_{ee}(\pm a_u-y)dy
\end{equation*}
or (for our particular weight functions) the formula
\begin{equation}
    \theta_u=2A_{ee}\sigma_{ee}e^{\frac{-a_u}{\sigma_{ee}}}\sinh\left(\frac{a_u}{\sigma_{ee}}\right).
\end{equation}
For such a solution to be self-consistent, we also must ensure that $V < \theta_v$ everywhere. Assuming the peak of the subthreshold I bump occurs at $x=0$, this is ensured by the condition
\begin{align*}
    V(0) = \int_{-a_u}^{a_u} w_{ie} (y) dy = 2A_{ie}\sigma_{ie}\left[1-e^{\frac{-a_u}{\sigma_{ie}}}\right] < \theta_v.
\end{align*}
There are thus, two branches of bump solutions. One branch of `Broad' bumps has superthreshold active regions in both the E and I populations ($a_u>0$ and $a_v>0$). For sufficiently high thresholds, we typically obtain marginally stable solutions, but these destabilize through a Hopf bifurcation for sufficiently low firing rate thresholds (See \cref{fig:2}A and subsequent stability analysis). In contrast, when the I bump is subthreshold ($V(0)< \theta_v$), we obtain unstable `Narrow' solutions that create a separatrix between the Broad solutions and the rest state (\cref{fig:2}A). Under each set of assumptions we obtain two distinct systems that approach the same solution defining a discontinuous saddle node bifurcation. The discontinuity arises from there being either two or one threshold condition, and the peak of the I Narrow bump grazing the threshold $\theta_v$. We derive these results in detail in the following subsection. 


\subsection{Eigenvalues and Noiseless Perturbations}
\begin{figure}[t!]
  \centering
  \includegraphics[width=\textwidth]{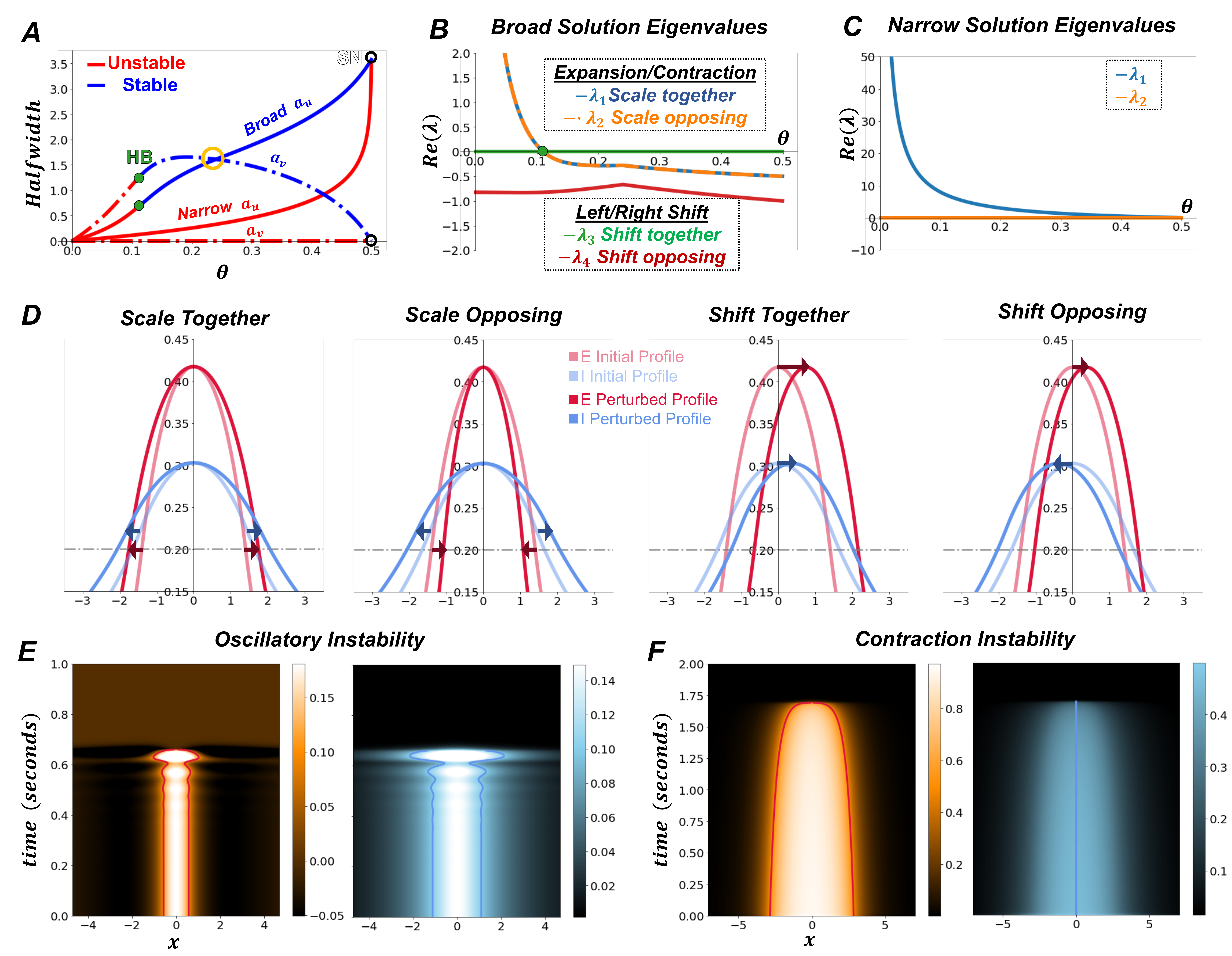}
  \caption{\textbf{Half-widths of bump solutions and linear stability.} (A) Broad/Narrow half-width solutions $a_u$ and $a_v$ as a function of firing rate threshold $\theta_u = \theta_v = \theta$. There is a Hopf bifurcation (HB, green dots) in the Broad solutions and a semi-stable discontinuous saddle node bifurcation (SN, white dots) where the Narrow and Broad solutions meet. The gold circle identifies where the E/I bump half-widths exchange size ordering, Eq.~\cref{eq:3.5}. 
  Eigenvalues are plotted in association with stability of the (B) Broad and (C) Narrow solutions. (D) Examples of four perturbations types related to the scale and shift Broad solution eigenvalues. (E) An example of an unstable Broad solution destabilized via the oscillatory instability emerging from the Hopf bifurcation, $\theta_u=\theta_v=0.1$. (F) A bump collapsing for threshold parameters near the discontinuous saddle node, $\theta_u=\theta_v=0.4996$. We take the bump-like solution initially narrower than the Broad solution. The I-I connectivity $A_{ii}=0$ in all. Other parameters are as in \cref{tableparam}.}
  \label{fig:2}
\end{figure}

Our stability analysis utilizes linearization about stationary solutions and localizes the perturbation evolution problem to the bump edges, as in several previous analyses of bump dynamics in neural fields~\cite{Blomquist05,coombes2003waves,folias2004breathing,Pinto2001b,Amari77}. 
Since there are four threshold crossing points, analysis of the Broad solution's stability results in four corresponding eigenvalues and equations associated with the degrees of freedom in the stability problem (\cref{fig:2}B). To derive our linearized system whose spectrum defines the stability of bumps, we start by perturbing with small smooth functions, $u(x,t)\approx U(x)+\epsilon \psi(x,t)$ and $v(x,t)\approx V(x)+\epsilon \phi(x,t)$. Substituting into Eq.~\cref{eq:3.1}, expanding about the stationary solution, and simplifying to first order we obtain the system:
\begin{subequations}
\begin{align}
    \psi_t + \psi &=w_{ee}(x)*[H'(U(x)) \psi(x,t)]-w_{ei}(x)*[H'(V(x))\phi(x,t)],\\
     \tau \phi_t + \phi &= w_{ie}(x)*[H'(U(x)) \psi(x,t)]-w_{ii}(x)*[H'(V(x))\phi(x,t)],
\end{align}\label{eq:3.6}
\end{subequations}
where the distributional derivatives are given
\begin{align*}
    H'(U(x))&=\frac{1}{|U'(a_u)|}(\delta(x-a_u)+\delta(x+a_u)), \\ H'(V(x))&=\frac{1}{|V'(a_v)|}(\delta(x-a_v)+\delta(x+a_v)).
\end{align*} 
Assuming separability of our perturbations, $\psi(x,t)=\psi(x)e^{\lambda t}$ and $\phi(x,t)=\phi(x)e^{\lambda t}$, and substituting our exponential weight functions and $H'(U(x))$ and $H'(V(x))$, we obtain the corresponding eigenvalue problem:
\begin{subequations} \label{evaleq}
\begin{equation}
\begin{split}
    (\lambda+1)\psi(x)=\frac{A_{ee}}{|U'(a_u)|}\left[e^{\frac{-|x-a_u|}{\sigma_{ee}}}\psi(a_u)+e^{\frac{-|x+a_u|}{\sigma_{ee}}}\psi(-a_u)\right]\\ -\frac{A_{ei}}{|V'(a_v)|}\left[e^{\frac{-|x-a_v|}{\sigma_{ei}}}\phi(a_v)+e^{\frac{-|x+a_v|}{\sigma_{ei}}}\phi(-a_v)\right],
\end{split}
\end{equation}
\begin{equation}
\begin{split}
    (\tau\lambda+1)\phi(x)=\frac{A_{ie}}{|U'(a_u)|}\left[e^{\frac{-|x-a_u|}{\sigma_{ie}}}\psi(a_u)+e^{\frac{-|x+a_u|}{\sigma_{ie}}}\psi(-a_u)\right]\\ -\frac{A_{ii}}{|V'(a_v)|}\left[e^{\frac{-|x-a_v|}{\sigma_{ii}}}\phi(a_v)+e^{\frac{-|x+a_v|}{\sigma_{ii}}}\phi(-a_v)\right].
\end{split}
\end{equation}
\end{subequations}
Note, in the above, we have assumed $\psi ( \pm a_u) \neq 0$ and $\phi ( \pm a_v) \neq 0$ to obtain an equation for the point spectrum, but taking these values to vanish would give us an equation for the essential spectrum which will not contribute to instabilities~\cite{coombes2004evans,veltz2010local}. We form a system of equations by evaluating the perturbation functions at the bump edges $x = \pm a_u$ and $x = \pm a_v$ into Eq.~(\ref{evaleq}). Values $\lambda$ for which the resulting $4 \times 4$ system is singular correspond to the four distinct eigenvalues in the point spectrum. Forming this system, taking the determinant, and factoring, we find the eigenvalues are the roots of the following pair of quadratics:
\begin{subequations}
\begin{align}
    \tau\lambda^2-(I+J+\tau B+ \tau C)\lambda+(I+J)(B+C)-(E+D)(F+G)\label{eq:3.8a}\\
    \tau\lambda^2-(I-J+\tau B- \tau C)\lambda+(I-J)(B-C)+(E-D)(F-G)\label{eq:3.8b}
\end{align}\label{eq:3.8}
\end{subequations}
where
\begin{alignat*}{2}
    & B=\frac{A_{ee}}{|U'(a_u)|}-1
    && \hspace{1in}C=\frac{A_{ee}}{|U'(a_u)|}e^{-\frac{2a_u}{\sigma_{ee}}}\\
    & D=-\frac{A_{ei}}{|V'(a_v)|}e^{\frac{-|a_v-a_u|}{\sigma_{ei}}}
    &&\hspace{1in} E=-\frac{A_{ei}}{|V'(a_v)|}e^{\frac{-|a_v+a_u|}{\sigma_{ei}}}\\
    & F=\frac{A_{ie}}{|U'(a_u)|}e^{\frac{-|a_v-a_u|}{\sigma_{ie}}}
    && \hspace{1in}G=\frac{A_{ie}}{|U'(a_u)|}e^{\frac{-|a_v+a_u|}{\sigma_{ie}}}\\
    & I=-\frac{A_{ii}}{|V'(a_v)|}-1
    && \hspace{1in}J=-\frac{A_{ii}}{|V'(a_v)|}e^{-\frac{2a_v}{\sigma_{ii}}}
\end{alignat*}
Each quadratic could also be obtained by restricting the form of perturbation (scaling and shifting) at the interfaces (See \cref{fig:2}D for examples). Scaling perturbations expand or contract the bump ($\psi(a_u)=\psi(-a_u)$ and  $\phi(a_v)=\phi(-a_v)$), and we obtain Eq.~\cref{eq:3.8a} resulting in the associated `scale' eigenvalues (\cref{fig:2}B):
\begin{equation}
    \lambda_{\text{scale}}=\frac{I+J+\tau (B+C)}{2\tau}\pm\frac{\sqrt{(I+J-\tau (B+C))^2+4\tau(E+D)(F+G)}}{2\tau},
\end{equation}
which are both nonzero in general. The sign of their real part determines the linear stability of the bumps. For shifting perturbations ($\psi(a_u)=-\psi(-a_u)$ and $\phi(a_v)=-\phi(-a_v)$) we obtain Eq.~\cref{eq:3.8b}. A straightforward calculation shows $(I-J)(B-C)+(E-D)(F-G)=0$, resulting in the zero eigenvalue (green) corresponding to the system's translational invariance (\cref{fig:2}B). Correlated shifts in the E/I bumps' center of mass simply translate the solution along the real line. Such perturbations are integrated, and neither grow nor decay. The nonzero shift eigenvalue is given by the simple formula
\begin{equation}
    \lambda=\frac{I-J+\tau(B-C)}{\tau}
\end{equation}
which is real and negative. Hence the bumps are stable with respect to shifts of the E and I bump in the opposite direction.

Eigenvalues of the Narrow solutions are obtained by following the same process, which is greatly simplified since the I population is subthreshold and perturbations of this part do not contribute to instabilities or enter into the point spectrum equations. We therefore investigate the effect of small smooth perturbations to the E population, with nonzero part along the bump edges, $u(x,t) \approx U(x) + \epsilon \psi (x,t)$. The resulting two eigenvalues are $\lambda_{1}=0$ (corresponding to the marginal stability of shifts) and $\lambda_2 = {2e^{\frac{-2a_u}{\sigma_{ee}}}}/{(1-e^{\frac{-2a_u}{\sigma_{ee}}})}$, which is clearly positive, confirming that Narrow solutions are always unstable (\cref{fig:2}C).

We identify bifurcations in the bump solutions by checking the signs of the real parts of each solution branch's eigenvalues. We find two types of bifurcations. The first is a Hopf bifurcation (green dot, HB in \cref{fig:2}A,B) occurring as the firing threshold is reduced, destabilizing bumps in a pattern-destroying oscillation (\cref{fig:2}E)~\cite{Pinto2001b}. This Hopf bifurcation boundary occurs when $Re(\lambda_{\text{scale}})=0$, which is given by the implicit formula 
\begin{equation}
    \frac{\tau A_{ee}}{|U'(a_u)|}(1-e^{\frac{-2a_u}{\sigma_{ee}}})=\tau+1+\frac{A_{ii}}{|V'(a_v)|}(1+e^{\frac{-2a_v}{\sigma_{ii}}}).
\end{equation}

The second bifurcation observed is a discontinuous saddle node bifurcation (SN in \cref{fig:2}A) \cite{Coombes2012nonsmooth,Leine2000} where the unstable Narrow and marginally stable Broad solutions meet. The peak of the Narrow I population bump rises to meet the firing threshold and the Broad I bump active region shrinks to a single point at threshold. The meeting of two linearizations of different discrete dimension (4 for Broad, 2 for Narrow) results in a discontinuous change in the corresponding Jacobian matrices yielding a semistable nonsmooth bifurcation where the solution branches annihilate one another (labelled ``contraction instability" in \cref{fig:2}F).

\subsection{Bounding stability regions in parameter space} 

\begin{figure}[t!]
  \centering
  \includegraphics[width=0.8\textwidth]{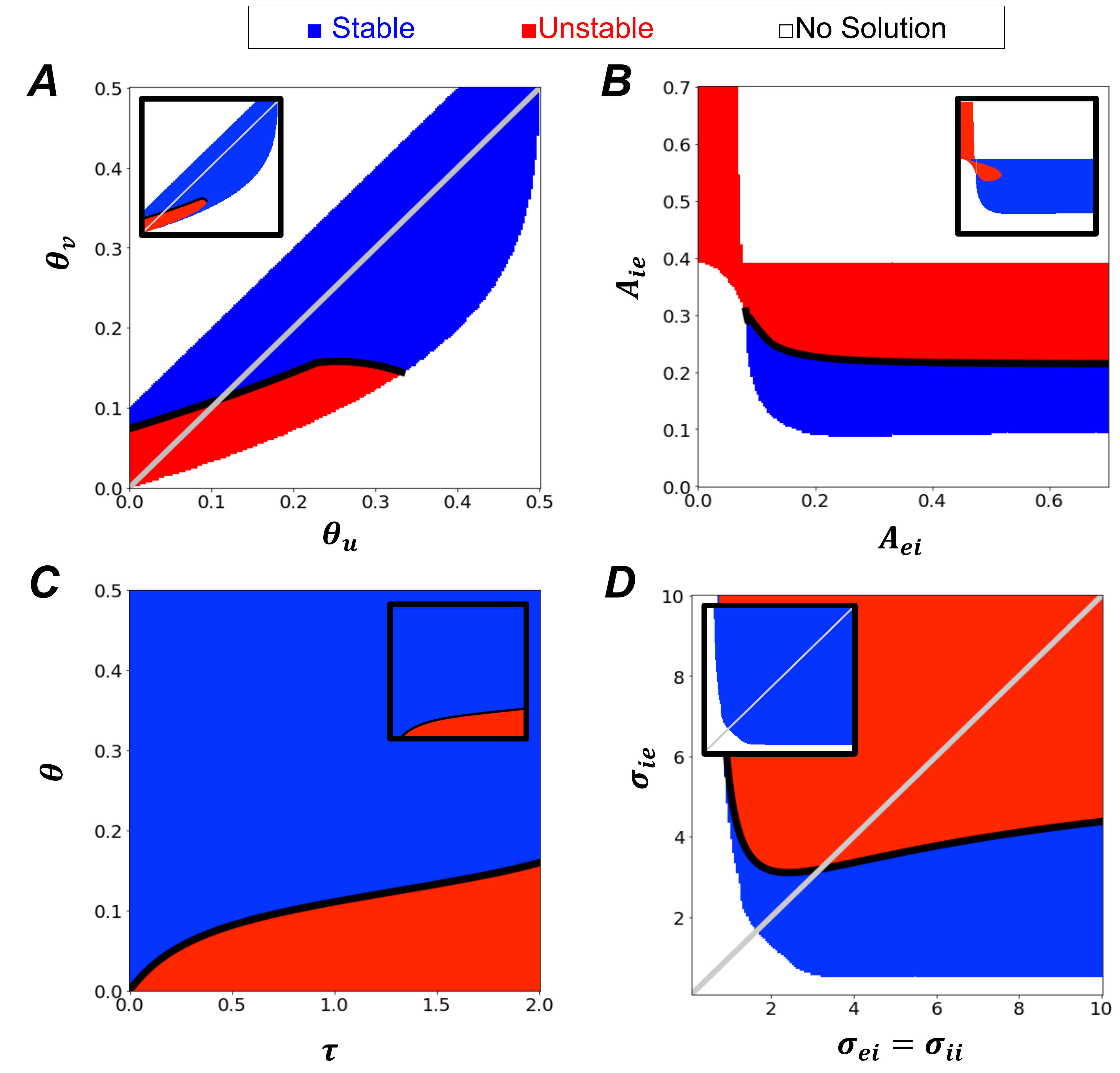}
  \caption{\textbf{Bump stability/instability across parameter space.} Main panels take $A_{ii}=0$ and insets take $A_{ii}=0.01$. In all plots, blue regions indicate parameters for which a stable bump exists, red regions are where all bumps are unstable. Black boundaries identify where Hopf bifurcations occur. White regions are where no bump solutions exist. (A) Stability as the firing rate thresholds $\theta$ are varied. 
  The silver line identifies where $\theta_u=\theta_v$. (B) Stability regions with varied E/I interaction amplitude $A_{ei}$ and $A_{ie}$. $\theta_u=\theta_v=0.25$.
  (C) Stability regions with varied inhibition timescale $\tau$ and firing thresholds $\theta_u = \theta_v = \theta$. 
  (D) Stability regions with varied E/I interaction spatial extent $\sigma_{ei} = \sigma_{ii}$ and $\sigma_{ie}$. $\theta_u=\theta_v=0.15$. 
  Silver line identifies where $\sigma_{ei}=\sigma_{ii}=\sigma_{ie}$. Other parameters are as in \cref{tableparam}.}  
  \label{fig:3}
\end{figure}

To identify how noise degrades stable representations of parametric working memory, we focus on the effect of stochastic perturbations to bumps along the Broad branch. Given that varying certain parameters can stabilize the deterministic system Eq.~\cref{eq:2.1}, we identify how the bifurcation boundaries change and stable solution regions expand/contract as parameters are varied (\cref{fig:3}). As discussed above, stable solutions only exist if Broad bumps have non-positive eigenvalues associated with their linear stability.

When varying the firing rate thresholds, we find larger I thresholds, $\theta_v$, expand the range of stable solutions (\cref{fig:3}A). We speculate that setting the I population firing rate threshold too low makes it strongly responsive to network activity perturbations, as shown in oscillatory instability simulations (\cref{fig:2}E).
Varying the strengths of synapses from E to I and I to E populations, we also observed bumps destabilize if E to I ($A_{ie}$) is  sufficiently strong (\cref{fig:3}B). Such changes increase the sensitivity of the E drive to the I population as bumps are expanded, leading to potential I bump overcompensation of as E bumps widen, eventually silencing both bumps.

Setting firing rate thresholds equal ($\theta_u = \theta_v = \theta$) and varying them (silver line in \cref{fig:3}A) along with changing the I timescale, $\tau$, results in a narrowing of the range of stable bumps as firing rate thresholds are decreased (\cref{fig:3}C). Specifically, as the timescale of the I population is increased, the system reacts more slowly to being out of equilibrium, so the E and I bump do not restabilize once perturbed. For instantaneous inhibition ($\tau \to 0$), such oscillatory (Hopf) instabilities never occur.

We also quantified the parameter ranges of stable bumps when varying the spatial scale of interpopulation synaptic footprints (\cref{fig:3}D). Stability is again largely dependent on the profile of E to I population synaptic connectivity; wider connectivity (high $\sigma_{ie}$) yields an unstable region. Similar to our previous findings, broadening the E to I profile leads to overcompensation of the I population in response to dynamical increases in the E population width.

The main panels of \cref{fig:3} assume $A_{ii}=0$, common in studies of Eq.~\cref{eq:2.1} due to the small amplitude of such connections. We  also examined small, nonzero amplitude I to I connections ($A_{ii} = 0.01$, See insets in \cref{fig:3}) and found stable regions expand likely due to dampening of I population reactions.

\section{Analysis of stochastic bump motion}
\label{sec:Stochastic}
Considering stochasticity emerging from neural and synaptic variability~\cite{faisal2008noise}, multiplicative noise in the neural field model Eq.~(\ref{eq:2.1}) (taking $\epsilon >0$) causes stable bumps to wander like a particle subject to Brownian motion. The extent of this wandering has been linked to subject response errors on delayed estimation tasks~\cite{Wimmer14,Compte2000}. We are primarily interested in identifying how architectural features of the E/I neural circuit impact wandering~\cite{Kilpatrick13WandBumpSIAD, Kilpatrick18STP_FacSCIREP, Seeholzer19,kilpatrick2013optimizing} as described by the time-dependent variance of bumps' center of mass. We can estimate this variance analytically using two different approaches.
The first we refer to as the `Strongly coupled limit' approximation, motivated by work from \cite{Kilpatrick13WandBumpSIAD}, which assumes the E/I bump has a single position (the E and I bump do not stray too far from one another). To first order, this approximation treats the bump as wandering by pure diffusion, and so its stochastic motion is only characterized by a diffusion coefficient. However, the resulting formulas have limitations for certain parameter regimes on the timescale of interest, since they do not consider that the bumps can drift apart. Our second approach, the `Interface-based approximation,' tracks bump interfaces (following threshold crossing points) \cite{Krishnan18,Coombes2012InterfaceJNeuro,Amari77}. Pairing this together with a weak coupling assumption as in \cite{Kilpatrick13FrontiersMultiArea} allows us to estimate the time-dependent changes in the distinct E and I bump centers of mass.

\subsection{Strongly coupled limit approximation}
In the Strongly coupled limit, we assume solutions to Eq.~\cref{eq:2.1} take the form of stable bumps both with position weakly perturbed by the same amount ($U(x-\Delta_u)$ and $V(x-\Delta_v)$) with $\Delta_u = \Delta_v = \Delta$. Thus, the E and I bumps are assumed to move together, and we have
\begin{subequations}
\begin{align}
    u(x,t)&=U(x-\Delta(t))+\epsilon^{1/2}\Phi(x-\Delta(t),t)+\epsilon\Phi_1(x-\Delta(t),t)\dots,\\
    v(x,t)&=V(x-\Delta(t))+\epsilon^{1/2}\Psi(x-\Delta(t),t)+\epsilon\Psi_1(x-\Delta(t),t)\dots,
\end{align}\label{eq:diffansatz}
\end{subequations}
where $\Delta(t)$ is the $\mathcal{O}(\epsilon^{1/2})$ position perturbation (with $d \Delta = \mathcal{O}(\epsilon^{1/2})$) and $\Phi$ and $\Psi$ are respectively the leading order profile perturbations to the E and I bump. Substituting Eq.~\cref{eq:diffansatz} and truncating Eq.~\cref{eq:2.1} to first order and taking averages, we find that the stationary solutions remain the same form as before. Moving to $\mathcal{O}(\epsilon^{1/2})$, we find
\begin{align}
   \left(\begin{array}{c}
     d\Phi\\
     \tau\cdot d\Psi
\end{array}\right)=\mathcal{L}
\left(\begin{array}{c}
     \Phi\\
     \Psi
\end{array}\right)dt+\epsilon^{-1/2} d\Delta(t)\left(\begin{array}{c}
     U'(x)\\
     \tau\cdot V'(x)
\end{array}\right)+\left(\begin{array}{c}
     \sqrt{|U(x)|}dW_u\\
     \sqrt{|V(x)|}dW_v
\end{array}\right),\label{eq:4.2}
\end{align}
where 
$$
\mathcal{L}\left(\begin{array}{c}
     p\\
     q
\end{array}\right)=
\left(\begin{array}{c}
     -p+w_{ee}*[f'(U)p]-w_{ei}*[f'(V)q]\\
     -q+w_{ie}*[f'(U)p]-w_{ii}*[f'(V)q]
\end{array}\right).
$$
Note that due to translation symmetry of the noise-free system, $\mathcal{N}(\mathcal{L})$ is spanned by $\{U',V'\}$. To briefly show this, recall the stationary solutions take the form 
\begin{align*}
    U(x)&=\int_{\mathbb{R}}w_{ee}(x-y)f(U(y))dy-\int_{\mathbb{R}}w_{ei}(x-y)f(V(y))dy\\
    V(x)&=\int_{\mathbb{R}}w_{ie}(x-y)f(U(y))dy-\int_{\mathbb{R}}w_{ii}(x-y)f(V(y))dy.
\end{align*}
Differentiating with respect to $x$ and applying integration by parts yields
\begin{align*}
    U'(x)&=\int_{\mathbb{R}}w_{ee}(x-y)f'(U(y))U'(y)dy-\int_{\mathbb{R}}w_{ei}(x-y)f'(V(y))V'(y)dy\\
    V'(x)&=\int_{\mathbb{R}}w_{ie}(x-y)f'(U(y))U'(y)dy-\int_{\mathbb{R}}w_{ii}(x-y)f'(V(y))V'(y)dy,
\end{align*}
exactly the form of the functions $p$ and $q$ spanning the $\mathcal{N}(\mathcal{L})$.

We ensure a bounded solution to Eq.~\cref{eq:4.2} by requiring the inhomogeneous part to be orthogonal to $\mathcal{N}(\mathcal{L^*})$. 
There exists a single vector spanning the ${\mathcal N}(\mathcal{L^*})$, denoted $(\varphi_1(x),\varphi_2(x))$. 
Enforcing our conditions for bounded solutions via requiring the inner product of the nullspace $(\varphi_1(x),\varphi_2(x))$ and inhomogeneity $(h_1(x), h_2(x))$ vanishes $\int_{\mathbb{R}} \varphi_1(x) h_1(x) + \varphi_2(x) h_2(x) d x = 0$, we isolate the bump position increment:
$$d\Delta(t)=-\epsilon^{1/2}\frac{\int_{\mathbb{R}}[\varphi_1(x)\sqrt{|U(x)|}dW_u(x,t)+\varphi_2(x)\sqrt{|V(x)|}dW_v(x,t)]dx}{\int_{\mathbb{R}}[\varphi_1(x)U'(x)+\tau V'(x)\varphi_2(x)]dx}.$$
Since the above is simply a weighted integral over the spatiotemporal noises, to first order, we have a Brownian stochastic differential equation (SDE) for the bump position.  
Given noise is white in time, we find that the mean over realizations is $\langle\Delta (t)\rangle=0$,
and the variance is:
\begin{align}\label{eq:4.3}
    \langle\Delta (t)^2\rangle=\epsilon \frac{\int_{\mathbb{R}}\varphi_1(x) \cdot \left[ C_u(x) * \varphi_1(x) +2 C_c(x)*\varphi_2(x) \right]+\varphi_2(x)\cdot \left[ C_v(x)*\varphi_2(x) \right]dx}
    {\left[\int_{\mathbb{R}}\varphi_1(x)U'(x)+\tau V'(x)\varphi_2(x)dx\right]^2}t
\end{align}
where the spatial correlation functions are defined:
$\langle W_u(x,t)W_u(y,t)\rangle=C_u(x-y)t$, 
$\langle W_v(x,t)W_v(y,t)\rangle=C_v(x-y)t$, and 
$\langle W_u(x,t)W_v(y,t)\rangle=C_c(x-y)t$. Note, there is a weighted contribution to the linear scaling variance from both the noise of the E and I populations.

To calculate the nullspace of the adjoint linearized operator, we rearrange the equation $\mathcal{L}^* \left( \varphi_1(x) , \varphi_2(x) \right)^T = 0$, so 
\begin{subequations}
\begin{align}
  \varphi_1(x)&=f'(U)\int_{\mathbb{R}}[w_{ee}(x-y)\varphi_1(y)+w_{ie}(x-y)\varphi_2(y)]dy,  \\
  \varphi_2(x)&=-f'(V)\int_{\mathbb{R}}[w_{ei}(x-y)\varphi_1(y)+w_{ii}(x-y)\varphi_2(y)]dy.
\end{align} \label{eq:4.4}
\end{subequations}
In the case of the Heaviside firing rate function we again can determine the distributional derivative of the Heavisde acting on the bump solution and obtain
\begin{align}
    \begin{pmatrix}
    \varphi_1(x)\\
    \varphi_2(x)
    \end{pmatrix}
    =
    \begin{pmatrix}
    \delta(x+a_u)+\mathcal{A}\delta(x-a_u)\\
    -\mathcal{B}\delta(x+a_v)+\mathcal{C}\delta(x-a_v)
    \end{pmatrix}.\label{eq:4.5}
\end{align}
Plugging Eq.~\cref{eq:4.5} into Eq.~\cref{eq:4.4} and solving for the constants, we find $\mathcal{A}=-1$, $\mathcal{C}=\mathcal{B}$, and 
$\mathcal{B}=\frac{w_{ei}(a_u-a_v)-w_{ei}(a_u+a_v)}{w_{ie}(a_u-a_v)-w_{ie}(a_u+a_v)}.$
Thus we find that 
\begin{align}
    \begin{pmatrix}
    \varphi_1(x)\\
    \varphi_2(x)
    \end{pmatrix}
    =
    \begin{pmatrix}
    \delta(x+a_u)-\delta(x-a_u)\\
    -\mathcal{B}[\delta(x+a_v)-\delta(x-a_v)]
    \end{pmatrix}.\label{eq:4.6}
\end{align}
Finally, substituting Eq.~\cref{eq:4.6} into Eq.~\cref{eq:4.3}, we have
\begin{align}
    \langle\Delta (t)^2\rangle&=\epsilon \frac{D_1-D_2+D_3}
    {2\left[|U'(a_u)|+\mathcal{B}\tau|V'(a_v)|\right]^2}t \label{eq:4.7}
\end{align}
where
\begin{align*}
    D_1&=\theta_u[C_u(0)-C_u(2a_u)],\\
    D_2&=2\mathcal{B}\sqrt{\theta_u\theta_v}[C_c(a_u-a_v)-C_c(a_u+a_v)],\\
    D_3&=\theta_v \mathcal{B}^2[C_v(0)-C_v(2a_v)].
\end{align*}
The contributions arising due to noise correlations within the E and I populations, $D_1$ and $D_3$, are positive. In contrast, the diffusion contribution from the cross-population correlation term $C_c$ is negative assuming the correlation function is monotone decreasing. Thus, noise correlations between the E and I populations reduce wandering. Here, we assume $C_c\equiv 0$. Later, we discuss the effects of nonzero correlated noise components across the E and I populations and compare these theoretical results to numerical simulations.

\subsection{Interface Based Approximation Theory}
\begin{figure}[h!]
  \centering
  \includegraphics[width=\textwidth]{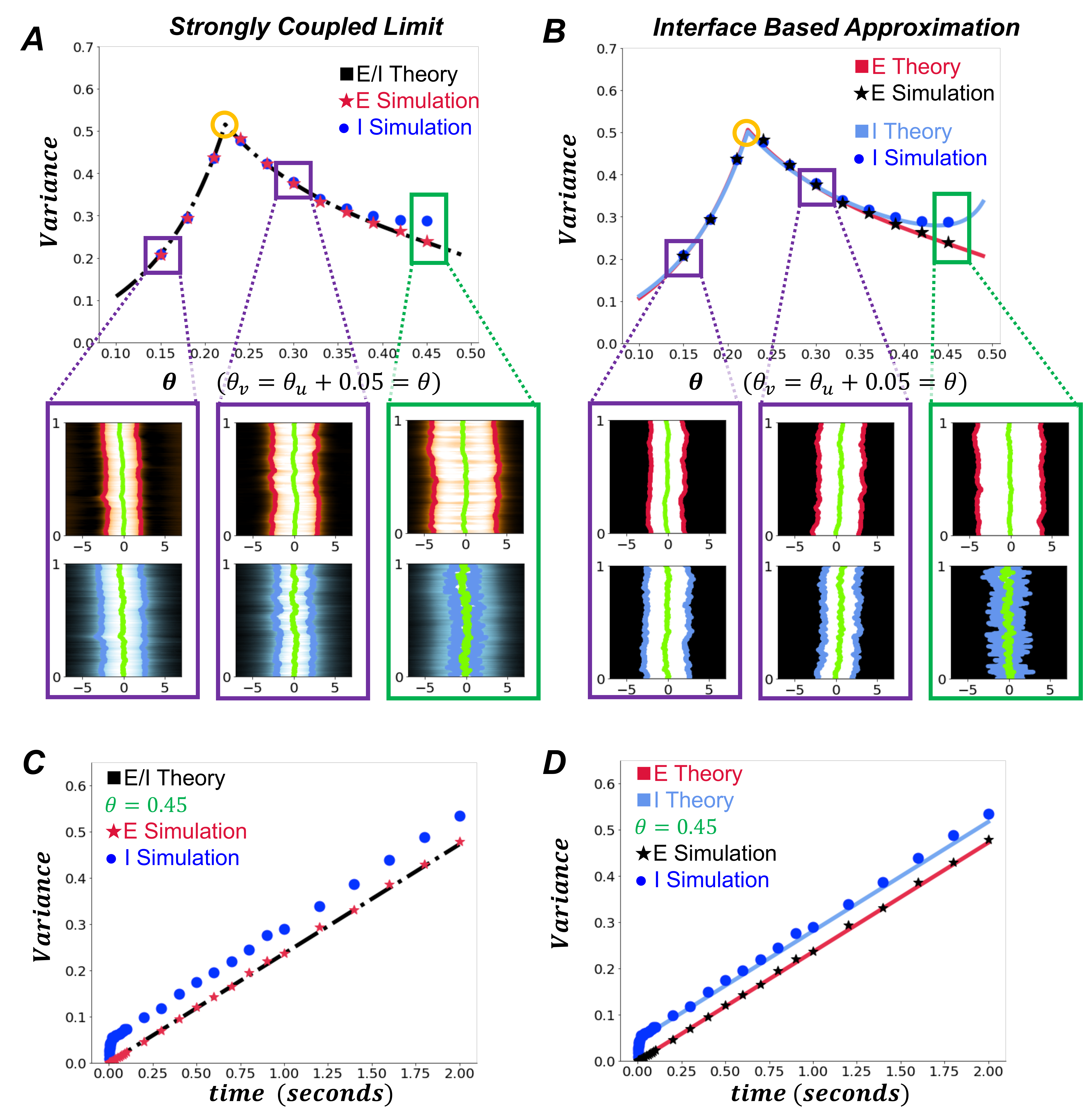}
  \caption{\textbf{Variance predictions and simulations.} Numerical simulations of Eq.~\cref{eq:2.1} were run using an approximate version of the interface equations derived from Eq.~\cref{eq:4.9}. Euler-Maruyama was used for time-stepping with noise amplitude $\epsilon=0.001$, the spatial interval was truncated to $[-3\pi,3\pi]$ with steps $dx=\frac{3\pi}{1000}$, timesteps are $dt = 1$ms, and variances were calculated by marginalizing over $10^4$ realizations per point. 
  (A)~The Strongly coupled limit prediction and corresponding simulations over 1 second. Although this prediction works reasonably well for small thresholds $\theta$, it breaks down for higher thresholds where the I bump center of mass differs considerably, e.g., $\theta_u+0.05=\theta_v=0.45$. Note the kink and change in the variance trend when $\theta$ passes through a value at which the half-widths $a_u$, $a_v$ exchange order (gold circle). Insets below show single E and I bump simulations at indicated threshold values. (B) When comparing to estimates of variance made using the Interface-based approach, theory more closely tracks the simulation results at higher firing rate threshold $\theta$. (C) The Strongly coupled limit predicts pure diffusion and linearly scaling variance, which underestimates variance calculated from simulations at $\theta_u+0.05=\theta_v=0.45$. (D) The Interface-based estimate tracks the drifting apart of the E and I bump, leading to more accurate variance predictions when $\theta_u+0.05=\theta_v=0.45$. All other parameters are as in \cref{tableparam}.
}
  \label{fig:4}
\end{figure}

Complementing our Strongly coupled limit approximation of bump wandering, we also derive interface equations that approximate the stochastic dynamics of the bump in response to multiplicative noise in Eq.~\cref{eq:2.1}. While the Strongly Coupled Limit approximation assumes the E and I bump move as one, numerical simulations reveal that this is not the case (See \cref{fig:4}A and C). For instance, the I bump tends to be more susceptible to profile deformations and may wander more, making the Strongly coupled limit less valid. Higher-order corrections of the variance estimate can be obtained using interface theory to develop nonlinearly coupled Langevin equations for estimating the E and I bumps' coupled centers of mass.

Interface theory for bumps in neural fields tracks the level sets of neural activity variables where firing rate thresholds are crossed. This approach was originally pioneered in the case of noiseless single-bump neural field models~\cite{Amari77} and then subsequently for traveling fronts in inhomogeneous neural fields~\cite{coombes2011pulsating} as well as solutions in planar neural fields~\cite{Coombes2012InterfaceJNeuro}. More recently, these approaches were adapted to obtain higher-order approximations for the timescale of front initiation~\cite{Faye2018} as well as the stochastic motion of multiple interacting bumps~\cite{Krishnan18}. Here, we further adapt this work \cite{Krishnan18} to account for the motion of separate bumps in the E and I neural populations of Eq.~\cref{eq:2.1}.

As defined previously, the active regions of the E and I bump are $[x_1(t),x_2(t)]$ and $[x_3(t),x_4(t)]$ respectively. Without noise perturbations, we would expect these interfaces to remain constant for the equilibrium bump solution, but noise perturbs these values so they wander over time. However, unlike in the Strongly coupled limit, we do not expect this wandering to be pure Brownian motion, but rather the result of a nonlinearly coupled system of SDEs. To obtain these SDEs, we start by writing the level set condition for the interfaces in each neural population (E and I):
\begin{align}
u(x_1(t),t)= u(x_2(t), t) =\theta_u \hspace{10mm} v(x_3(t),t)=v(x_4(t),t)=\theta_v, \label{eq:4.8}
\end{align}
where $\theta_u$ and $\theta_v$ are the firing rate thresholds of the E and I populations. For a Heaviside firing rate, we substitute Eq.~\cref{eq:4.8} into Eq.~\cref{eq:2.1} and obtain
\begin{subequations}
\begin{align}
    du&=\left[-u+\int^{x_2(t)}_{x_1(t)}w_{ee}(x-y)dy-\int^{x_4(t)}_{x_3(t)}w_{ei}(x-y)dy\right]dt+\epsilon^{\frac{1}{2}}dW_{e}\\
    \tau dv&=\left[-v+\int^{x_2(t)}_{x_1(t)}w_{ie}(x-y)dy-\int^{x_4(t)}_{x_3(t)}w_{ii}(x-y)dy\right]dt+\epsilon^{\frac{1}{2}}dW_{i}
\end{align}\label{eq:4.9}
\end{subequations}
Differentiating Eq.~\cref{eq:4.8}, we obtain consistency equations for the interfaces $x_j(t)$:
\begin{align*}
    \partial_x u(x_j(t),t) d x_j + d u(x_j(t), t) &= 0, \ \ j=1,2; \\ \partial_x v(x_j(t),t) d x_j + d v(x_j(t), t) &= 0, \ \ j = 3,4.
\end{align*}
We then approximate the above exact evolution equations by assuming the spatial gradients at the interfaces remain constant and odd symmetric throughout the evolution motivated by the form of the Strongly coupled limit expansion:
\begin{align*}
|U'(a_u)| &\equiv \alpha_u(t)\approx \frac{\partial u(x_1(t),t)}{\partial x}=-\frac{\partial u(x_2(t),t)}{\partial x}, \\
|V'(a_v)| &\equiv \alpha_v(t) \approx \frac{\partial v(x_3(t),t)}{\partial x}=-\frac{\partial v(x_4(t),t)}{\partial x}.
\end{align*}
Subsequently, we drop $\mathcal{O}(\epsilon)$ terms to obtain the relations
\begin{subequations}
\begin{align}
    du(x_{1}(t),t)&=-\alpha_{u}dx_1(t),\\
    du(x_{2}(t),t)&=\alpha_{u}dx_2(t),\\
    dv(x_{3}(t),t)&=-\alpha_{v}dx_3(t),\\
    dv(x_{4}(t),t)&=\alpha_{v}dx_4(t).
\end{align}\label{eq:4.10}
\end{subequations}
Plugging the formulas in Eq.~\cref{eq:4.10} into \cref{eq:4.9}, we obtain the following system of nonlinear Langevin equations describing the stochastic evolution of the interfaces:
\begin{subequations}
\begin{align}
    dx_1&=-\frac{1}{\alpha_u}\left(
    \left[-\theta_u+\mathcal{W}_{ee}(x_1; x_1, x_2)-\mathcal{W}_{ei}(x_1;x_3,x_4)
    \right]dt +\epsilon^{\frac{1}{2}}dW_e(x_1,t)\right)\\
    dx_2&=\frac{1}{\alpha_u}\left(
    \left[-\theta_u+\mathcal{W}_{ee}(x_2; x_1, x_2)-\mathcal{W}_{ei}(x_2; x_3, x_4)
    \right]dt +\epsilon^{\frac{1}{2}}dW_e(x_2,t)\right)\\
    \tau dx_3&=-\frac{1}{\alpha_v}\left(
    \left[-\theta_v+\mathcal{W}_{ie}(x_3;x_1,x_2)-\mathcal{W}_{ii}(x_3;x_3,x_4)
    \right]dt +\epsilon^{\frac{1}{2}}dW_i(x_3,t)\right)\\
    \tau dx_4&=\frac{1}{\alpha_v}\left(
    \left[-\theta_v+\mathcal{W}_{ie}(x_4;x_1,x_2)-\mathcal{W}_{ii}(x_4;x_3,x_4)
    \right]dt +\epsilon^{\frac{1}{2}}dW_i(x_4,t)\right),
\end{align}\label{eq:4.11}
\end{subequations}
where the coupling functions are given by the integrals over the active regions
\begin{align*}
    \mathcal{W}_{jk}(x;x_a, x_b) = \int^{x_b}_{x_a}w_{jk}(x-y)dy.
\end{align*}
To derive estimates for the evolution of the centers of mass of the E and I bump, we apply the definitions $\Delta_u = (x_1 + x_2)/2$ and $\Delta_v = (x_3+x_4)/2$ from Eq.~\cref{eq:2.3} to Eq.~\cref{eq:4.11} and combine equations to obtain
\begin{subequations}
\begin{align}
\begin{split}
     d\Delta_u=\frac{1}{2\alpha_u}\left(\left[
    -2\mathcal{W}_{ee}(x_1;x_1,x_2)+\mathcal{W}_{ei}(x_1;x_3,x_4)-\mathcal{W}_{ei}(x_2;x_3,x_4)
    \right]dt\right.\\\left.+\sqrt{\epsilon \theta_u}[dW_u(x_2,t)-dW_u(x_1,t)]
    \right),
\end{split}
\\
\begin{split}
     d\Delta_v=\frac{1}{2\tau\alpha_u}\left(\left[
    2\mathcal{W}_{ii}(x_3;x_3,x_4)+\mathcal{W}_{ie}(x_4;x_1,x_2)-\mathcal{W}_{ei}(x_3;x_1,x_2)
    \right]dt\right.\\\left.+\sqrt{\epsilon \theta_v}[dW_v(x_4,t)-dW_v(x_3,t)]
    \right).
\end{split}
\end{align}\label{eq:4.12}
\end{subequations}
Assuming position perturbations remain small we approximate $x_{1}=\Delta_u- a_u$, $x_{2}=\Delta_u+ a_u$, $x_{3}=\Delta_v- a_v$, and $x_{4}=\Delta_v+ a_v$ with $\Delta_u, \Delta_v = \mathcal{O}(\epsilon^{1/2})$. Plugging in these approximations, linearizing about the stationary solutions, and simplifying yields the multivariate Ornstein-Uhlenbeck (OU) process:
\begin{subequations}
\begin{align}
    \begin{split}
     d\Delta_u(t)=\frac{1}{\alpha_u}\left((\Delta_u-\Delta_v) \cdot \left[w_{ei} (a_u -a_v) - w_{ei}(a_u+a_v) \right]dt\right.\\\left.+\sqrt{\epsilon\theta_u}[dW_u(\Delta_u+a_u,t)-dW_u(\Delta_u-a_u,t)]\right)
\end{split}
\\
\begin{split}
     \tau d\Delta_v(t)=\frac{1}{\alpha_v}\left((\Delta_v-\Delta_u) \cdot \left[ w_{ie}(a_u + a_v)- w_{ie}(a_u-a_v) \right]dt\right.\\\left.+\sqrt{\epsilon\theta_v}[dW_v(\Delta_v+a_v,t)-dW_v(\Delta_v-a_v,t)]\right).
\end{split}
\end{align}
\end{subequations}
Note, $w_{ei}$ and $w_{ie}$ are even and monotonic in distance, so we define the positive quantaties
\begin{align*}
M_u&=\alpha_u^{-1} \cdot \left[ w_{ei}(a_u - a_v) - w_{ei}(a_u+a_v) \right]>0, \\
M_v&=\tau^{-1} \alpha_v^{-1} \cdot \left[ w_{ie}(a_u - a_v)- w_{ie}(a_u+a_v) \right]>0.
\end{align*}
Then we have the matrix-vector representation of the multivariate OU process,  
$d\boldsymbol{\Delta}=\boldsymbol{\mathcal{K}}\boldsymbol{\Delta}dt+\boldsymbol{d\mathcal{W}}$,
where $\boldsymbol{\Delta}=\begin{pmatrix}
\Delta_u\\ \Delta_v
\end{pmatrix}$,  $\boldsymbol{\mathcal{K}}=\begin{pmatrix}
M_u&-M_u\\M_v&-M_v
\end{pmatrix}$ is the coupling matrix, and $$\boldsymbol{d\mathcal{W}}=\begin{pmatrix}
\frac{\sqrt{\epsilon\theta_u}}{2\alpha_u}[dW_u(\Delta_u+a_u,t)-dW_u(\Delta_u-a_u,t)]\\ \frac{\sqrt{\epsilon\theta_v}}{2\tau\alpha_v}[dW_v(\Delta_v+a_v,t)-dW_v(\Delta_v-a_v,t)]
\end{pmatrix}$$
is the correlated Wiener process noise. Following methods for solving linear SDEs~\cite{Gardiner2009}, we can determine the mean and variance of $\Delta_u$ and $\Delta_v$ to obtain estimates of the diffusion of each population.
First we diagonalize $\boldsymbol{\mathcal{K}}$:
$$\boldsymbol{\mathcal{K}}=\frac{1}{M_v-M_u}\begin{pmatrix}
1&M_u\\1&M_v
\end{pmatrix}\begin{pmatrix}
0&0\\0&-(M_v-M_u)
\end{pmatrix}\begin{pmatrix}
-M_v&M_u\\-1&1
\end{pmatrix}$$
Thus our eigenvalues are $\lambda_{1,2}=0,-(M_v-M_u)$ and the eigenvectors are $\mathbf{v}_{1}=(1,1)^T$ and $\mathbf{v}_{2}=(M_u,M_v)^T$ respectively. Note, this implies that there is a marginally stable direction along perturbations of the bump that move both the E and I bump the same amount, and there is an attractive (stable) direction for perturbations that move the E and I bump differently. Similar results have been found for coupled lateral I layers for which each layer individually supports a self-sustaining bump in the absence of cross-population coupling~\cite{FoliasErmentrout2011,Kilpatrick13FrontiersMultiArea,bressloff2015nonlinear}.
The mean is given by $\langle\boldsymbol{\Delta}\rangle=e^{\boldsymbol{\mathcal{K}}t}\boldsymbol{\Delta}(0)$, and so
\begin{align*}
    \begin{pmatrix}
    \langle\Delta_u\rangle\\
    \langle\Delta_v\rangle
    \end{pmatrix}=\frac{1}{M_v-M_u}
    \begin{pmatrix}
    M_v\Delta_v(0)-M_u\Delta_u(0)-M_u[\Delta_u(0)-\Delta_v(0)]e^{-(M_v-M_u)t}\\
    M_v\Delta_v(0)-M_u\Delta_u(0)+M_v[\Delta_v(0)-\Delta_u(0)]e^{-(M_v-M_u)t}
    \end{pmatrix}
\end{align*}
Next we seek the covariance $\langle\boldsymbol{\Delta}(t)\boldsymbol{\Delta}^T(t)\rangle=\int^t_0e^{\boldsymbol{\mathcal{K}}(t-s)}\boldsymbol{D}e^{\boldsymbol{\mathcal{K}^T}(t-s)}ds$ where the covariance matrix of the noise term, $\boldsymbol{D}$ is found to be:
\begin{align}
    \boldsymbol{D}=\begin{pmatrix}
    D_u&D_c\\
    D_c& D_v
    \end{pmatrix}
\end{align}
with
\begin{align*}
    &D_u=\frac{\epsilon \theta_u}{2\alpha_u^2}[C_u(0)-C_u(2a_u)]\\
    &D_v=\frac{\epsilon \theta_v}{2\tau^2\alpha_v^2}[C_v(0)-C_v(2a_v)]\\
    \begin{split}
        D_c=\frac{\epsilon \sqrt{\theta_u\theta_v}}{4\tau\alpha_u\alpha_v}[C_c(\Delta_u-\Delta_v+a_u-a_v)-C_c(\Delta_u-\Delta_v+a_u+a_v)\\-C_c(\Delta_u-\Delta_v-a_u-a_v)+C_c(\Delta_u-\Delta_v-a_u+a_v)].
    \end{split}
\end{align*}
Multiplying out $e^{\boldsymbol{\mathcal{K}}(t-s)}\boldsymbol{D}e^{\boldsymbol{\mathcal{K}^T}(t-s)}$ and then integrating yields our predictions of E and I bump center of mass variance:
\begin{subequations} \label{eq:distinct}
\begin{align}
    \begin{split}
        \langle\Delta_u(t)^2\rangle&=\frac{D_vM_u^2-2D_cM_uM_v+D_uM_v^2}{(M_v-M_u)^2}t\\
        &-2\frac{e^{-(M_v-M_u)t}-1}{(M_v-M_u)^3}[D_cM_u^2-D_vM_u^2+D_cM_uM_v-D_uM_uM_v]\\
        &-M_u^2\frac{e^{-2(M_v-M_u)t}-1}{2(M_v-M_u)^3}[D_u+D_v-2D_c],
    \end{split}
    \\
    \begin{split}
        \langle\Delta_v(t)^2\rangle&=\frac{D_vM_u^2-2D_cM_uM_v+D_uM_v^2}{(M_v-M_u)^2}t\\
        &-2\frac{e^{-(M_v-M_u)t}-1}{(M_v-M_u)^3}[D_cM_v^2-D_uM_v^2+D_cM_uM_v-D_vM_uM_v]\\
        &-M_v^2\frac{e^{-2(M_v-M_u)t}-1}{2(M_v-M_u)^3}[D_u+D_v-2D_c].
    \end{split}
\end{align}
\end{subequations}
In the limit as $t\rightarrow \infty$ we find that both variances are dominated by the term:
\begin{align*}
        \langle\Delta_u(t)^2\rangle=\langle\Delta_v(t)^2\rangle=\frac{D_vM_u^2-2D_cM_uM_v+D_uM_v^2}{(M_v-M_u)^2}t,
\end{align*}
which is essentially an estimate of the variance derived from assuming the E and I bump are co-located as in the Strongly coupled limit approximation. Relating these two routes to one another, we find
$$\mathcal{B}=\frac{w_{ei}(a_u-a_v)-w_{ei}(a_u+a_v)}{w_{ie}(a_u-a_v)-w_{ie}(a_u+a_v)}=\frac{\alpha_u M_u}{\tau\alpha_v M_v}.$$
Using this relation and plugging in the expressions for $D_u$, $D_v$, and $D_c$ we find that we obtain the diffusion coefficient expression from the Strongly coupled limit prediction, Eq.~\cref{eq:4.7}. At short times, the Interface-based approximation, Eq.~\cref{eq:distinct}, has contributions based on the interactions of the E and I bumps, which decay over time.

The main difference between the Strongly coupled limit and Interface-based methods is that the bumps are allowed to drift apart in the Interface-based method, which allows us to separately estimate the I bump's variance. The most general form of the approximation can further describe stochastic widening and contraction of bump, described by a full four-dimensional and nonlinear approximation. Even collapsing to centers of mass approximations, using the Interface-based interactions as a starting point is more accurate than the Strongly coupled limit, as we will subsequently show. On the other hand, the Strongly coupled limit is more straightforward to obtain, and there are fewer arbitrary assumptions we must make (e.g., the static gradient approximation). Yet, the Interface-based method allows us to obtain greater precision by using the fully nonlinear approximation over all the interfaces, $x_j$, making it a better metric for variance predictions. 

\subsection{Variance Predictions and Simulations Comparison}

To validate our Strongly coupled limit and Interface-based predictions of bump variance, we ran stochastic simulations of Eq.~\cref{eq:2.1} using spatiotemporal noises to the E and I populations that are not correlated between populations; $C_c(x)\equiv 0$. 
As firing rate thresholds are increased (\cref{fig:4}A-B), both our theoretical predictions and the averaged numerical simulations suggest that variance changes non-monotonically. This stands in stark contrast to results from previous studies, which found that the effective diffusion of bumps generally tends to increase monotonically with the firing threshold for single-population lateral I networks~\cite{Kilpatrick13WandBumpSIAD}. Interestingly, we find that the variance peaks at the precise point in parameter space where the noise-free E and I bump have the same width (gold circles, \cref{fig:4}A,B).

Moreover, we see that the Strongly coupled limit approximation adequately captures the effective motion of the E bump, but not the I bump, which tends to stray further (\cref{fig:4}A). In contrast, the Interface-based approximation is able to capture both the distinct I and E bumps' variance by accounting for the stochastic dynamics of the I bump being perturbed away from the E bump center of mass (\cref{fig:4}B). 

The dynamics of the I bump vary with firing rate thresholds and timescales. At higher firing rate thresholds, the I bump wanders more than the E bump, and the two bumps tend to be more weakly coupled. As the timescale increases, the I bump center of mass relaxes to wander slightly away from that of the E bump, while the E bump variance scales linearly in time like pure diffusion (\cref{fig:4}C,D). Since the I bump is sustained by input from the E population, the I bump appears to weakly track the input(see \cref{fig:EIWANDER} for single simulation of such activity), which can be estimated by OU processes~\cite{Kilpatrick13WandBumpSIAD,bressloff2015nonlinear}. Hence, the I bump variance can be higher than that of the E population.

\begin{figure}[t!]
  \centering
  \includegraphics[width=\textwidth]{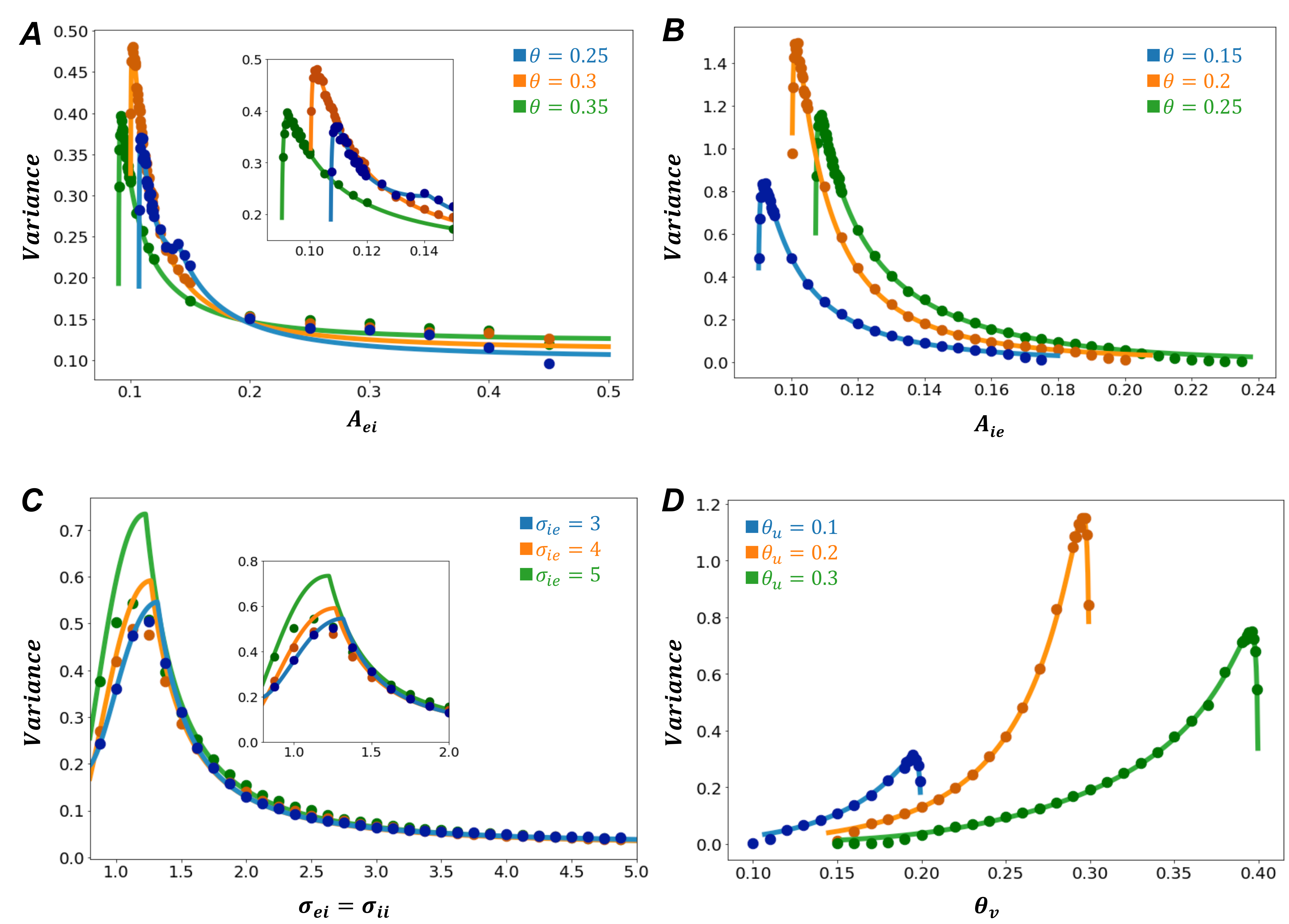}
  \caption{\textbf{Excitatory bump position variances as a function of network connectivity, spatial extent, and firing rate threshold.} Noise amplitude $\epsilon=0.001$ throughout. (A) Bump position variance mostly decreases as  connectivity amplitude $A_{ie}$ is increased. Other parameters are $ A_{ii}=0, \tau=1$, and $\theta_u=\theta_v=\theta$. (B) Bump position variance mostly decreases as $A_{ei}$ is increased. Other parameters are $A_{ii}=0, \tau=1$. $\theta_u=\theta_v=\theta$. (C) Bump position variance changes non-monotonically as the spatial extent of the I projections ($\sigma_{ei}=\sigma_{ii}$) are increased. Other parameters are $A_{ii}=0, \theta_u=\theta_v=0.3$ and $\tau=1$. $\sigma_{ie}$ is varied vary. (D) Bump positive variance primarily increases with I population threshold $\theta_v$. Other parameters are $ A_{ii}=0$ and $\tau=1$. We also vary firing threshold $\theta_u$. Other parameters are as in \cref{tableparam}. 
}
  \label{fig:5}
\end{figure}

To further analyze how the variance in bump position depends on changes in parameters, we determined how it changes along several different parameter axes (\cref{fig:5}). For simplicity, we studied the variations in the E bump's position variance and compared to the Interface-based approximation. We started by varying the amplitude of inter-population connectivity, either $E \to I$ or $I \to E$. Weakening either of these projection amplitudes tended to increase bump variance (\cref{fig:5}A,B). We speculate that weakening cross-population connectivity leads to less input between the bumps. Thus, the bumps are more weakly stable to noise perturbations and re-equilibrate more slowly, ultimately leading to more wandering.

Aside from non-monotonicity arising as a function of the firing rate threshold, we also observe that narrowing the I bump synaptic profile (varying the spatial extent of the I projections) yields a peak in variance (\cref{fig:5}C) which may be due to a greater susceptibility to noise perturbations when the width of I to E projections $\sigma_{ei} = \sigma_{ii}$ is decreased, resulting in a peak of variance. For smaller $\sigma_{ei} = \sigma_{ii}$, we speculate that the equilibrium E bump width is narrower and can be more easily stabilized by I feedback. Note that the simulations do not match the peaks as well as in other panels likely due to approximations made in the Interface-based approach, though the trends are still largely captured.

We also observe that lower I firing rate thresholds lead to bumps that are more stabilized to noise perturbations (\cref{fig:5}D).
However, the largest peaks in bump position variance occur where E and I bump half-widths are most similar, i.e., where E (I) threshold crossing gradients are the lowest (highest) (\cref{fig:supp3A}). 

These results assumed I to I connections are non-existent $A_{ii}=0$. However, we found that even low amounts of connectivity,  $I\rightarrow I$ connectivity $A_{ii}=0.01$, decreased bump variance (\cref{fig:supp5Aiinonzero}).

\begin{figure}[htbp]
  \centering
  \includegraphics[width=\textwidth]{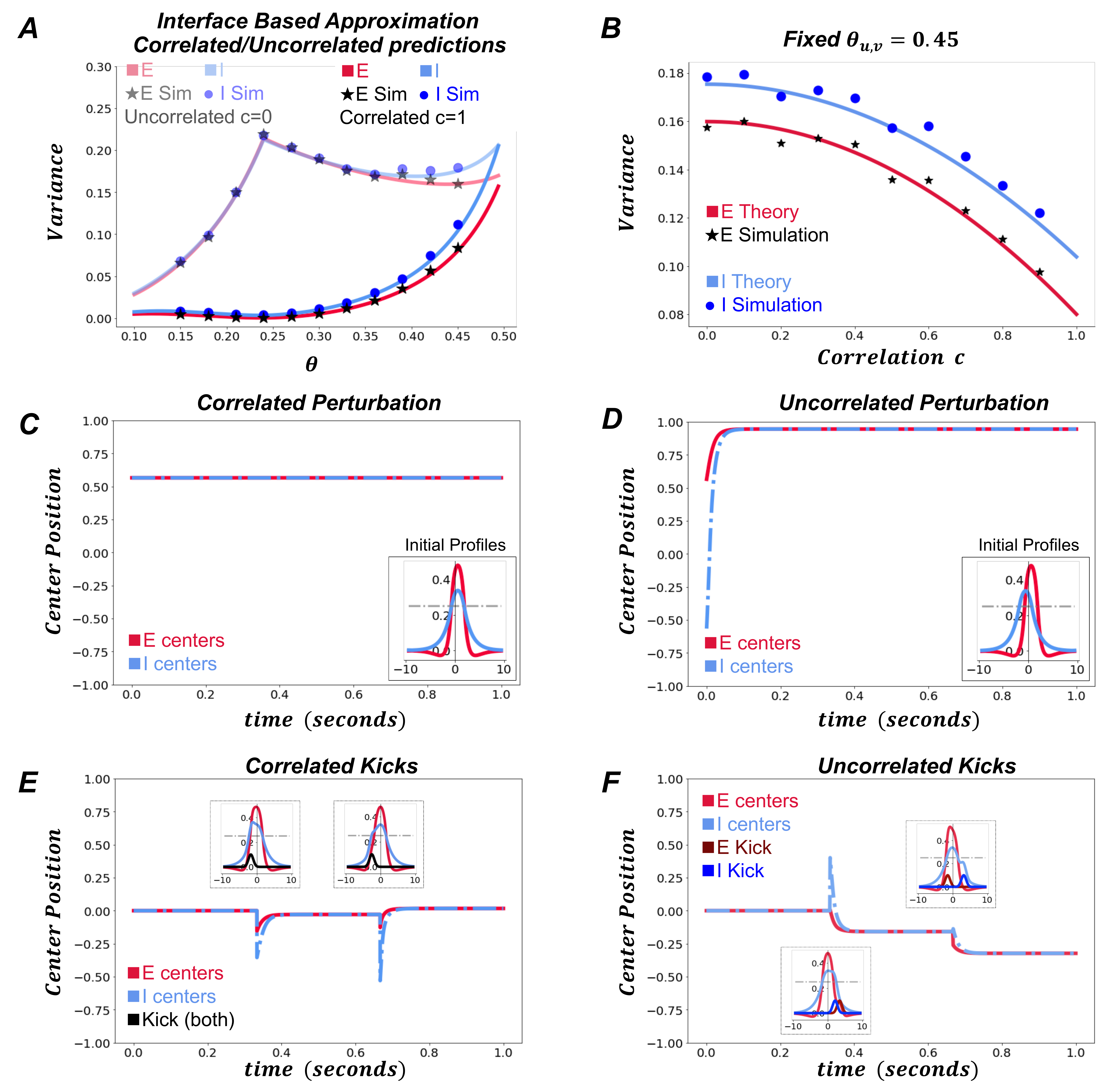}
  \caption{\textbf{Effects of inter-population noise correlations.} (A) Bump position variance as a function of firing rate threshold $\theta_u=\theta_v=\theta$ for a network with completely uncorrelated ($c=0$) or completely correlated ($c=1$) inter-population noise correlations. Interface-based theory (solid) agrees well with numerical simulations. (B) Bump position variance for fixed firing rate threshold $\theta_u=\theta_v=0.45$ as a function of inter-population noise correlations $c$. (C) Bump profile evolution in absence of noise for firing rate threshold $\theta_u=\theta_v=0.25$ given a correlated center shift perturbation at $t=0$. (D) Bump profile evolution in absence of noise over 1 second for firing rate threshold $\theta_u=\theta_v=0.25$ given an uncorrelated center shift perturbation at $t=0$. (E) Bump profile evolution for firing rather threshold $\theta_u=\theta_v=0.25$ given two correlated `kicks' (0.1 amplitude Gaussian bumps shifted by random amount) applied to each population. (F) Bump profile evolution for firing rate threshold $\theta_u=\theta_v=0.25$ with two uncorrelated `kicks' (0.1 amplitude Gaussian bumps shifted by random amount) applied to each population. Other parameters are $A_{ii} = 0$, $\epsilon = 0.001$ and otherwise as in \cref{tableparam}.
}
  \label{fig:6}
\end{figure}

\subsection{Correlated versus uncorrelated noise}
The impact of noise correlations on neural circuit codes for delayed estimates are varied. Correlated noise within a neural subpopulation can improve working memory coding \cite{leavitt2017correlated}, but cross-population noise correlations can lead to an increase in bump attractor wandering that degrades memory~\cite{Kilpatrick13FrontiersMultiArea}. We find in the subsequent investigation that cross-population noise correlations between E and I populations lead to less bump wandering than uncorrelated noise. 

Our aim is to explore the system when there is cross-population noise (corresponding to the case where $C_c\neq 0$). To obtain greater control over the extent of correlated noise we opt to express our spatiotemporal noise sources in the E and I populations as
\begin{equation}
    dW_{u,v}(x,t)\rightarrow \sqrt{1-c^2}dW_{u,v}(x,t)+cdW_c(x,t),
\end{equation}
where the independent spatially correlated and temporally white noise process $dW_c$ represents a correlated stochastic component in system Eq.~\cref{eq:2.1}. We define these three noise terms the same as before though now we have nonzero cross population spatial correlation $C_{c}(x-y)t=\langle W_c(x,t)W_c(y,t)\rangle$. The correlation parameter $c\in [0,1]$ such that $c=0$ implies uncorrelated noise, and $c=1$ fully correlated noise. Correlating noise across the E and I populations drastically alters to our variance predictions (\cref{fig:6}A and \cref{fig:supp6}), with increased correlated noise decreasing the predicted and simulated bump variances (\cref{fig:6}B). 

We analyze the response of the E/I bump structure to correlated as opposed to uncorrelated shifts. Correlated shifts translated both bumps centers, due to translational invariance of the system (\cref{fig:6}C). Uncorrelated shifts move the E and I bumps in opposite directions, but then show an attraction of the I bump to the E bump, while the E bump is repulsed by the I bump leading the I bump to `catch up' to the shifted E bump and the E bump moves further away from its original position (\cref{fig:6}D). Thus, uncorrelated shifts lead to additional drift of the bump and higher variance. This stands in stark contrast to the uncorrelated/correlated perturbation analysis carried out for two coupled identical lateral I layers~\cite{Kilpatrick13FrontiersMultiArea}, in which opposing perturbations of two weakly connected bumps are effectively canceled by the attractive force between the bumps. 

We also considered the effects of two small additive Gaussian inputs, more akin to the noisy kicks arising in stochastic simulations. When the position of these kicks was strongly correlated (\cref{fig:6}E; i.e. the same application to each population), the E and I bumps stayed together and relaxed back to their original position. Uncorrelated kicks led to different perturbations in the position and profile of the E and I bumps (e.g., \cref{fig:6}F), with a relaxation where the I bump was attracted the E bump, but the E bump was repelled by the I bump.

Overall, we found that the separate E/I network model shows parametrically-dependent bump wandering. First, bump position variance depends strongly on the relative widths of the E and I bumps, and relaxation dynamics from noise perturbations can cause the I bump to stray from the E bump. Second, non-monotonic variance trends arise with respect to network connectivity amplitude and spatial scale parameters. Lastly, increases in inter-population noise correlations reduce bump wandering by eliminating the relaxation effects that would otherwise extend the effects of stochastic perturbations on the E and I bumps.

\section{Discussion}
\label{sec:conclusions}

Stochastic bump attractor models have become a useful tool for characterizing variability of systems that code for memory-based estimates of continuous variables~\cite{Compte2000,Wimmer14,barbosa2020interplay,kim2017ring}. As we show here, it is important to appreciate the nuances of E/I architecture and noise-correlations~\cite{GoldmanRakic1995} when making predictions about continuous variable estimates. Our study has provided a suite of new predictions concerning how inter-population connectivity amplitude and spatial profiles impact bump wandering. To obtain tractable expressions for bump position variance predictions, one form of our approximations relied upon the marginal stability of solutions to the noise-free system. Prior to performing our variance estimates, we observed that along several parameter axes, we obtain non-monotonic changes in half-widths and two types of instabilities: an oscillatory (Hopf) instability located in the red unstable regions, and a contraction instability located at semistable discontinuous saddle nodes (\cref{fig:3}). Partial versions of the results have been observed previously in E/I population models~\cite{Blomquist05}. We also found that stability of solutions was significantly affected by nonzero I$\rightarrow$I synaptic strength (compare \cref{fig:3} and insets). Thus, even `weak' I$\rightarrow$I connectivity can strongly affect the linearized dynamics of stationary bump solutions.

Our asymptotic analysis aimed at predicting the stochastic motion of bumps subject to noise moved beyond the standard single center-of-mass approximation often used to estimate the stochastic motion of bumps~\cite{Laing2001,Compte2000,Kilpatrick13WandBumpSIAD}. While a single center-of-mass approximation works reasonably well across some parts of parameter space, at high firing rate threshold, close to the discontinuous saddle-node bifurcation, we find this breaks down and is better captured by a stochastic interface based approximation previously developed in \cite{Krishnan18}. In particular, the I bump diffuses more than the E bump, which is well captured by the nonlinear Langevin approximation derived by approximating the motion of bump interfaces. The amplitude of bump wandering changes non-monotonically in most network parameters, due to the change in bump half-width amplitudes, well captured by our Interface-based approximation. Notably, inter-population noise correlations reduced bump wandering. Uncoordinated E and I bump motion, arising in networks with uncorrelated inter-population noise, lead to additional bump drift during relaxation periods while the I bump `chased' the E bump. 

There are several possible extensions of our analysis of bump stochastic motion in the E/I network. It is important to note that we made multiple approximations to collapse our stochastically evolving bump interface equations to a pair of SDEs describing the coupling between the E and I bump. Alternatively, we could have retained a higher order approximation of the bump interface gradients in order to obtain a more accurate approximation~\cite{Faye2018}. Recall, the Interface Based approximation begins as a fully nonlinear description and can be used to describe dynamics near the oscillatory (Hopf) bifurcation or the discontinuous saddle node. Alternatively, such near-bifurcation approximations could also be determined by choosing a scaling for a bifurcation parameter similar to the weak noise amplitude as in \cite{kilpatrick2014pulse,kilpatrick2016ghosts}.

Our analysis of bump position variance across multiple parametric axes helped identify a number of ways to reduce bump wandering via network architectural tuning. We largely chose parameters roughly assuming $80\%$ E and $20\%$ I neurons present in the prefrontal cortex~\cite{abeles1991corticonics}, but we could certainly explore broader ranges of parameter space beyond this typical fraction. Another natural extension for this work would be to consider more complex mechanisms for synaptic tuning, such as short term plasticity in the E and I populations, to reduce the propensity for bumps to wander. Recent studies in mean field reductions of spiking networks have demonstrated that short term facilitation (depression) on the E population with global inhibition tends to decrease (increase) bump drift and diffusion~\cite{Seeholzer19}, ultimately improving parametric working memory. Extensions of our model and present analysis could be used to further investigate how the introduction of different forms of short term plasticity into either the E or I population would impact bump wandering. Not only could short term plasticity reduce the effect of stochastic perturbations on bumps, but also make them more robust to distraction inputs~ \cite{Lorenc21}.

Finally, our investigation into the role of cross-population noise correlations raises questions regarding the coding advantages brought about by noise correlations. Noise correlations can increase, decrease, or not affect the amount of information encoded by a neural circuit~\cite{averbeck2006neural}. Most such results have been derived in network models devoid of spatial structure. However, recently work has demonstrated how disruptive broadly correlated spatiotemporal noise can be to information transmission in spatially-organized neural circuits~\cite{huang2020internally}. Our work adds to this ongoing line of inquiry by demonstrating variability-reducing mechanisms possible via increased correlation in noise between E and I populations. Our bump position variance predictions and model are sensitive to changes in the structure of noise. An improved understanding of the precise form and structure of noise in prefrontal cortex and other areas~\cite{mcdonnell2011benefits} could help further constrain neural circuit models of memory-encoding persistent activity. Mechanistic models  that connect synaptic architecture, psychophysical performance, stochastic and spatiotemporal dynamics, as well as the rich structure of internal and external noise can help us further understand the dynamical principles underlying information coding in the brain.

\section*{Acknowledgments}
We thank Sage Shaw for assistance with python code development.

\bibliographystyle{siamplain}

\appendix
\renewcommand\thefigure{\thesection.\arabic{figure}}   
\renewcommand\thetable{\thesection.\arabic{table}}  
\section{Additional Figures}

\setcounter{figure}{0}    
\begin{figure}[htbp]
  \centering
  \includegraphics[width=\textwidth]{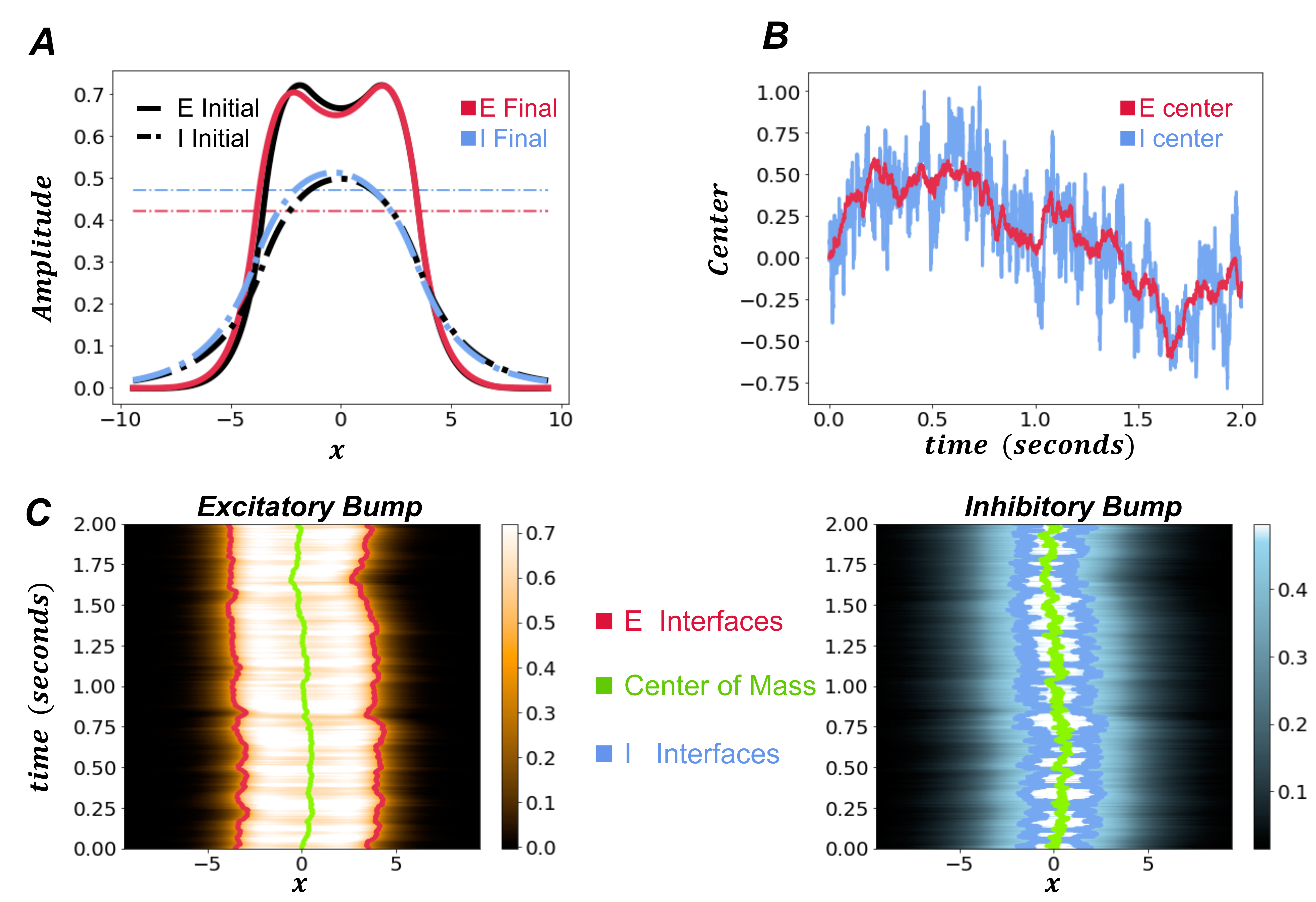}
  \caption{\textbf{Single simulation of E and I bump wandering.} The noise amplitude is $\epsilon=0.001$. The I bump wanders much further in certain parameter regimes. Synaptic weight strength $A_{ii}=0$; and firing rate thresholds $\theta_u=0.42$, and  $\theta_v=0.47$.  (A) Initial and final profiles of E and I bumps. (B) E and I center of mass positions evolving over time. The I bump generally follows E wandering, but also additionally wanders more wildly about E. (C) Bumps wandering over time. Traces represent the center of mass (green) and the E/I interfaces (red/blue, respectively). Colorscales represent the profile amplitude of $u(x,t)$ and $v(x,t)$. Other parameters are as in \cref{tableparam}.
}
  \label{fig:EIWANDER}
\end{figure}

\begin{figure}[htbp]
  \centering
  \includegraphics[width=0.8\textwidth]{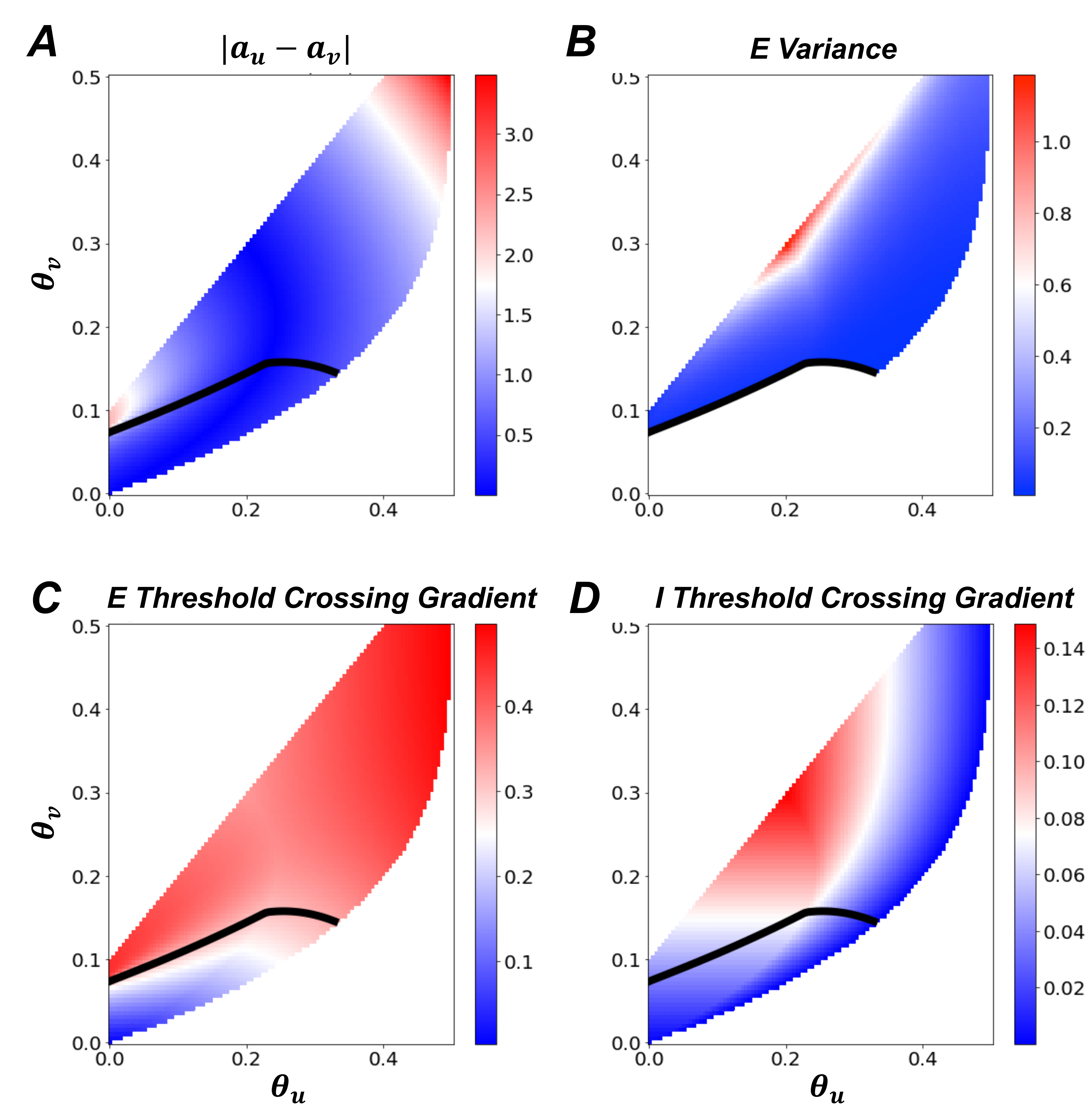}
  \caption{\textbf{Isolating factors affecting variance of bumps' wandering.} Synaptic strength $A_{ii}=0.01$. (A) Colorscale represents the difference in relative E and I halfwidths for varying $\theta_u$ and $\theta_v$. Bump position variances are maximized when $a_u=a_v$. (B) Colorscale of E bump position variance for $\epsilon=0.001$ amplitude noise. Black line bounds unstable region. (C,D) E,I bump threshold crossing gradient ($|U'(a_u)|$ and $|V'(a_v)|$) as a function of firing rate thresholds. Note, the E variance is largest when the threshold crossing gradient for E (I) is low (high). Slices further taken through these plots correspond to predictions shown in \cref{fig:5}D.
}
  \label{fig:supp3A}
\end{figure}

\begin{figure}[htbp]
  \centering
  \includegraphics[width=\textwidth]{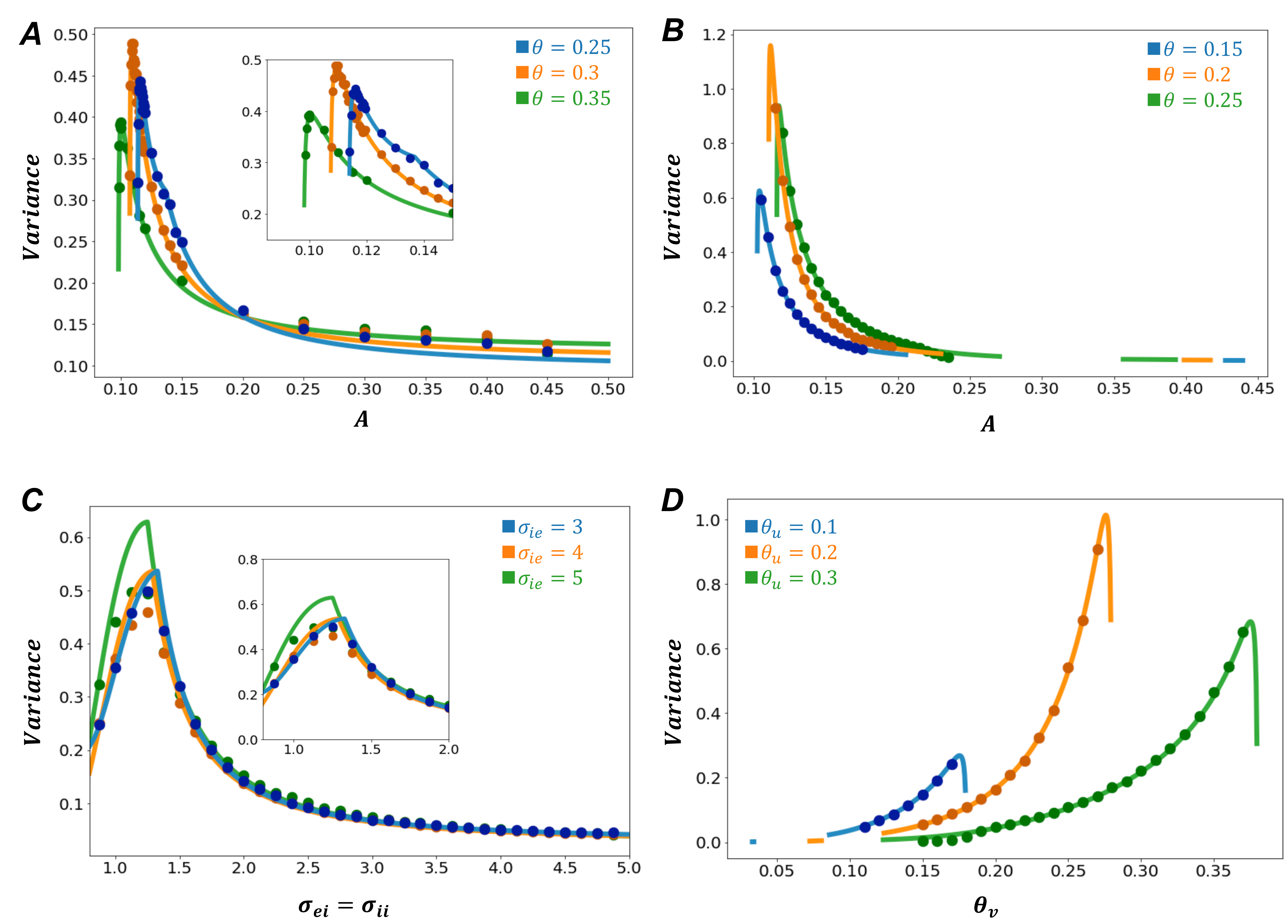}
  \caption{\textbf{Predicted and simulated center of mass variances.} Here, we take I to I connectivity $A_{ii}=0.01$ and noise amplitude $\epsilon=0.001$. (A) Bump position variance as a function of $A_{ei}=A$ vary and $\theta_u=\theta_v=\theta$ are varied. (B) Variance as a function of $A_{ie}=A$ as $\theta_u=\theta_v=\theta$ are also varied. (C) Variance as a function of spatial scale $\sigma_{ei}=\sigma_{ii}$ and varying $\sigma_{ie}$. We fix $\theta_u=\theta_v=0.3$. (D) Variance as a function of $\theta_v$ as $\theta_u$ is varied. Parameters not mentioned are as defined in \cref{tableparam}.
}
  \label{fig:supp5Aiinonzero}
\end{figure}

\begin{figure}[htbp]
  \centering
  \includegraphics[width=\textwidth]{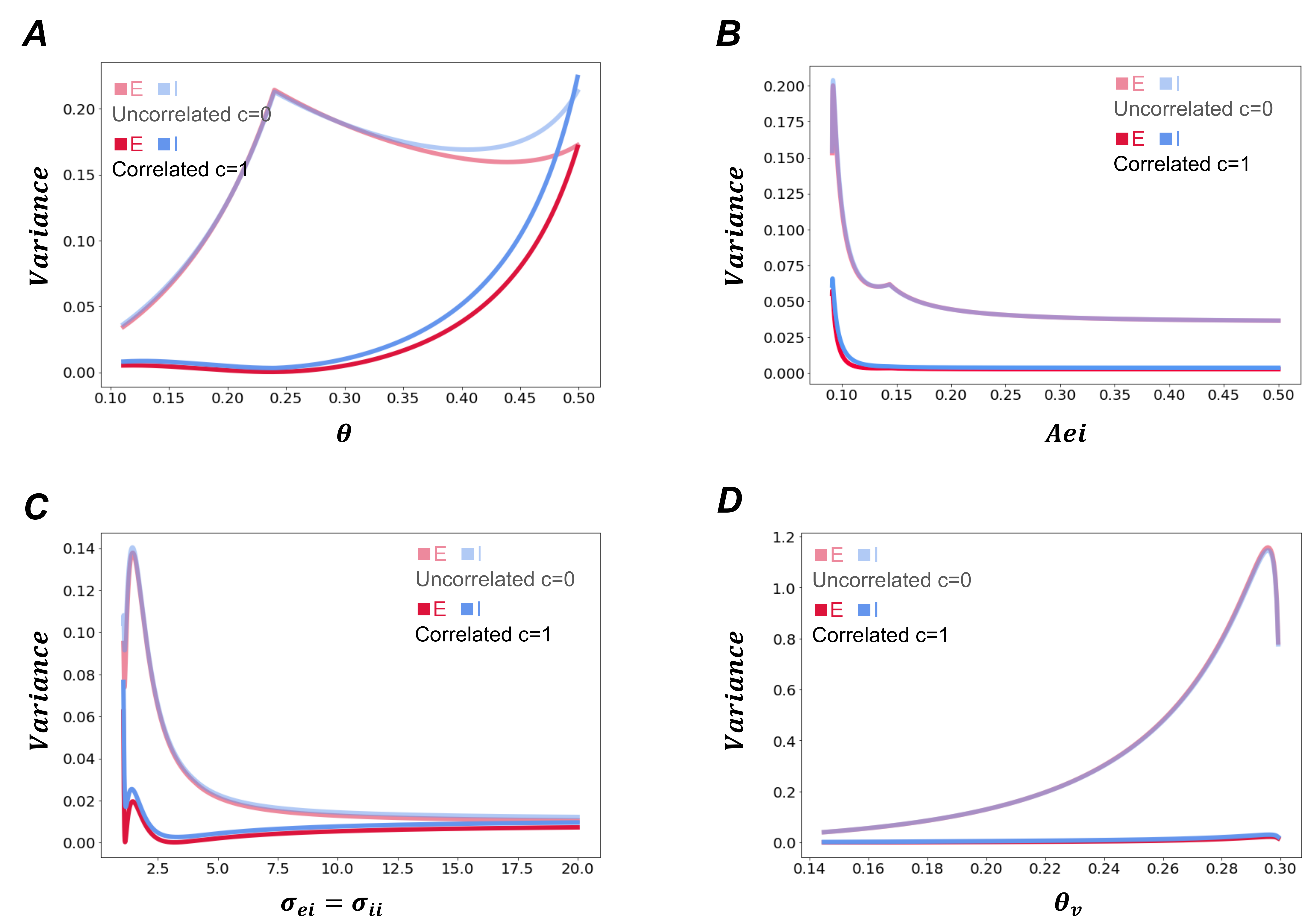}
  \caption{\textbf{Predicted and simulated bump position variances for correlated and uncorrelated noise.} Noise amplitude is set to $\epsilon=0.001$ and we take $A_{ii}=0$. (A)Bump position variance as a function of $\theta_u=\theta_v=\theta$.  (B) Bump position variance as a function $A_{ei}=A$ with $\theta_u=\theta_v=0.25$. (C) $\theta_u=\theta_v=0.2$ and $\sigma_{ie}=3$. Bump position variance as a function of $\sigma_{ei}=\sigma_{ii}$. (D) Bump position variance as a function of $\theta_v$. We take $\theta_u = 0.2$. Other parameters are as in \cref{tableparam}.
}
  \label{fig:supp6}
\end{figure}

\end{document}